\documentclass[10pt,letterpaper]{article}
\usepackage[top=0.85in,footskip=0.75in,left = 0.75in, right = 0.75in]{geometry}
\usepackage{amsmath,amssymb}
\usepackage{float}
\usepackage{changepage}
\usepackage[utf8x]{inputenc}
\usepackage{textcomp,marvosym}
\usepackage{cite}
\usepackage{nameref,hyperref}
\usepackage{microtype}
\usepackage{url}
\DisableLigatures[f]{encoding = *, family = * }
\usepackage[table]{xcolor}
\usepackage{array}
\usepackage{bibref}
\usepackage{dsfont}
\usepackage{cancel}
\usepackage{caption}
\usepackage{tikz-cd}
\usepackage{mathtools}
\usepackage{titlesec}
\titleformat{\paragraph}
{\normalfont\normalsize\bfseries}{\theparagraph}{1em}{}
\titlespacing*{\paragraph}
{0pt}{3.25ex plus 1ex minus .2ex}{1.5ex plus .2ex}

\newcolumntype{+}{!{\vrule width 2pt}}
\newlength\savedwidth

\usepackage[aboveskip=1pt,labelfont=bf,labelsep=period,justification=raggedright,singlelinecheck=off]{caption}

\makeatletter
\renewcommand{\@biblabel}[1]{\quad#1.}
\makeatother

\usepackage{lastpage,fancyhdr,graphicx}
\usepackage{epstopdf}
\begin{document}

\begin{flushleft}
{\Large
\textbf\newline{The microscopic relationships between triangular arbitrage and cross-currency correlations in a simple agent based model of foreign exchange markets} 
}
\newline
\\
Alberto Ciacci\textsuperscript{1,2,$\star$},
Takumi Sueshige\textsuperscript{3,$\star$},
Hideki Takayasu\textsuperscript{4,5},
Kim Christensen\textsuperscript{1,2},
Misako Takayasu\textsuperscript{3,4,*}
\\
\bigskip
\textbf{1} Blackett Laboratory, Imperial College London, London SW7 2AZ, United Kingdom
\\~\\
\textbf{2} Center for Complexity Science, Imperial College London, London SW7 2AZ, United Kingdom
\\~\\
\textbf{3} Department of Mathematical and Computing Science,
School of Computing, Tokyo Institute of Technology, 4259-G3-52, 
Nagatsuta-cho, Midori-ku, Yokohama 226-8503, Japan, 226-8502
\\~\\
\textbf{4} Institute of Innovative Research, Tokyo Institute of Technology 4259, 
Nagatsuta-cho, Yokohama 226-8502, Japan
\\~\\
\textbf{5} Sony Computer Science Laboratories, 3-14-13, Higashigotanda, Shinagawa-ku, Tokyo 141-0022, Japan
\\
\bigskip

%
%

$\star$ These authors contributed equally to this work \\~\\
* Corresponding author\\~\\
E-mail: takayasu.m.aa@m.titech.ac.jp
\end{flushleft}
\section*{Abstract}
Foreign exchange rates movements exhibit significant cross-correlations even on very short time-scales. 
The effect of these statistical relationships become evident during extreme market events, such as flash crashes.
In this scenario, an abrupt price swing occurring on a given market is immediately followed by anomalous movements 
in several related foreign exchange rates. 
Although a deep understanding of cross-currency correlations would be clearly beneficial for conceiving more stable and safer foreign exchange markets, the microscopic origins of these interdependencies have not been extensively investigated.
We introduce an agent-based model which describes the emergence of cross-currency correlations 
from the interactions between market makers and an arbitrager. 
Our model qualitatively replicates the time-scale vs. cross-correlation diagrams observed in real trading data, 
suggesting that triangular arbitrage plays a primary role in the entanglement of the dynamics of different foreign exchange rates. 
Furthermore, the model shows how the features of the cross-correlation function between two foreign exchange rates, such as its sign and value, emerge from the interplay between triangular arbitrage and trend-following strategies.
\section{Introduction}
Various non-trivial statistical regularities, known as \emph{stylized facts} \cite{buchanan2011s}, have been documented in trading data from markets of different asset classes \cite{gould2013limit}. The available literature examined the heavy-tailed distribution of price changes \cite{gopikrishnan1998inverse, cont2001empirical,plerou2008stock,chakraborti2011econophysics}, the long memory in the absolute mid-price changes (volatility clustering) \cite{liu1997correlations,cont2001empirical,cont2007volatility,stanley2008statistical, gu2009emergence,chakraborti2011econophysics}, the long memory in the direction of the order flow \cite{zovko2002power,lillo2004long,bouchaud2004fluctuations,gu2009emergence} and the absence of significant autocorrelation in mid-price returns time series, with the exclusion of negative, weak but still significant autocorrelation observed on extremely short time-scales \cite{cont2005long, stanley2008statistical,zhao2010model,ait2011ultra,chakraborti2011econophysics}.
Different research communities (e.g., physics, economics, information theory) took up the open-ended challenge of devising models that could reproduce these regularities and provide insights on their origins \cite{farmer2005economics, parlour2008limit, gould2013limit}. Economists have traditionally dealt with optimal decision-making problems in which perfectly rational agents implement trading strategies to maximize their individual utility
\cite{farmer2005economics, parlour2008limit, gould2013limit}. Previous studies have looked at cut-off decisions \cite{chakravarty1995integrated,parlour1998price,foucault1999order}, asymmetric information and fundamental prices \cite{glosten1985bid,kyle1985continuous,goettler2006microstructure,rocsu2009dynamic,rosu2010liquidity} and price impact of trades \cite{bertsimas1998optimal,almgren2001optimal,alfonsi2010optimal,obizhaeva2013optimal}.
In the last thirty years the orthodox assumptions of full rationality and perfect markets have been increasingly disputed by emerging disciplines, such as behavioral economics, statistics and artificial intelligence \cite{farmer2005economics}. The physics community have also entered this \emph{quest for simple models of non-rational choice} \cite{farmer2005economics} by taking viewpoints and approaches, such as zero-intelligence and agent-based models, that often stray from those that are common among economists. Agent-based models (ABMs henceforth) rely on simulations of interactions between agents whose actions are driven by idealized human behaviors \cite{farmer2005economics}. A seminal attempt to describe agents interactions through ABM is the \emph{Santa Fe Stock Market} \cite{anderson1988economy}, which neglects the perfect rationality assumption by taking an artificial intelligence approach \cite{farmer2005economics}. The model successfully replicates various stylized facts of financial markets (e.g., heavy-tailed distribution of returns and volatility clustering), hinting that the lack of full rationality has a primary role in the emergence of these statistical regularities \cite{farmer2005economics}. Following \cite{anderson1988economy}, several 
ABMs \cite{takayasu1992statistical, sato1998dynamic, cont2000herd, chiarella2002simulation, challet2003non,aiba2006microscopic, preis2006multi, preis2007statistical, lillo2007limit, yamada2007characterization, yamada2009solvable,lee2015heterogeneous, cocco2017using} have further examined the relationships between the microscopic interactions between agents and the macroscopic behavior of financial markets.\\ 
In this study we introduce a new ABM of the foreign exchange market (FX henceforth). This market is characterized by singular institutional features, such as the absence of a central exchange, exceptionally large traded volumes and a declining, yet significant dealer-centric nature \cite{rime2013anatomy}. 
Electronic trading has rapidly emerged as a key channel through which investors can access liquidity in the FX market \cite{rime2013anatomy, BIS}. For instance, more than 70\% of the volume in the FX Spot market is exchanged electronically \cite{BIS}. A peculiar stylized fact of the FX market is the significant correlation among movements of different currency prices. These interdependencies are time-scale dependent \cite{mizuno2004time,wang2013statistical}, their strength evolves in time and become extremely evident in the occurrence of extreme price swings, known as flash crashes. In these events, various foreign exchange rates related to a certain currency abruptly appreciate or depreciate, affecting the trading activity of several FX markets. A recent example is the large and rapid appreciation of the Japanese Yen against multiple currencies on January $2^{\text{nd}}$ 2019. The largest intraday price changes peaked +11\% against Australian Dollar, +8\% against Turkish Lira and +4\% against US Dollar \cite{han2019anatomy}. 
The relationship between triangular arbitrage \cite{aiba2002triangular,fenn2009mirage,kozhan2012execution,foucault2016toxic} and cross-currency correlations remains unclear. Mizuno \textit{et al.} \cite{mizuno2004time} observed that the cross-correlation between real and implied prices of Japanese Yen is significantly below the unit on very short time-scales, conjecturing that this counter-intuitive property highlights how the same currency could be purchased and sold at different prices by implementing a triangular arbitrage strategy. Aiba and Hatano \cite{aiba2006microscopic} proposed an ABM relying on the intriguing idea that triangular arbitrage influences the price dynamics in different currency markets. However, this study fails to explain whether and how reactions to triangular arbitrage opportunities lead to the characteristic shape of the time-scale vs cross-correlation diagrams observed in real trading data \cite{mizuno2004time, wang2013statistical}. 

Building on these observations, the present study aims to obtain further insights on the microscopic origins of the correlations among currency pairs. We introduce an ABM model in which two species (i.e., market makers and the arbitrager) interact across three inter-dealer markets where trading is organized in limit order books. The model qualitatively replicates the characteristic shape of the cross-correlation functions between currency pairs observed in real trading data. This suggests that triangular arbitrage is a pivotal microscopic mechanism behind the formation of cross-currency interdependencies. Furthermore, the model elucidates how the features of these statistical relationships, such as the sign and value of the time-scale vs cross-correlation diagram, stem from the interplay between trend-following and triangular arbitrage strategies.\\
This paper is organized as follows. Section \ref{sec:methods} presents the methods of this study. In particular, we outline the basic concepts, discuss the employed dataset and provide a detailed description of the proposed model. In Section \ref{sec:results}, we examine the behavior of our model against real FX markets. Section \ref{sec:final} concludes and provides an outlook on the research paths that could be developed from the outcomes of this study. Technical details, further empirical analyses and an extended version of the model
are presented in the supporting information sections.
\section{Methods}\label{sec:methods}
\subsection{Concepts}
\subsubsection{Limit Order Books}
Electronic trading takes place in an online platform where traders submit buy and sell orders for a certain assets through an online computer program. Unmatched orders \emph{await} for execution in electronic records known as limit order books (LOBs henceforth). By submitting an order, traders pledge to sell (buy) up to a certain quantity of a given asset for a price that is greater (less) or equal than its limit price \cite{gould2013limit,bonart2018continuous}. The submission activates a trade-matching algorithm which determines whether the order can be immediately matched against earlier orders that are still queued in the LOB \cite{bonart2018continuous}. A matching occurs anytime a buy (sell) order includes a price that is greater (less) or equal than the one included in a sell (buy) order. When this occurs, the owners of the matched orders engage in a transaction. Orders that are completely matched upon entering into the system are called \emph{market orders}. Conversely, orders that are partially matched or completely unmatched upon entering into the system (i.e., \emph{limit orders}) are queued in the LOB until they are completely matched by forthcoming orders or deleted by their owners \cite{bonart2018continuous}.  
\begin{figure}[H]
\centering
\includegraphics[scale=0.15]{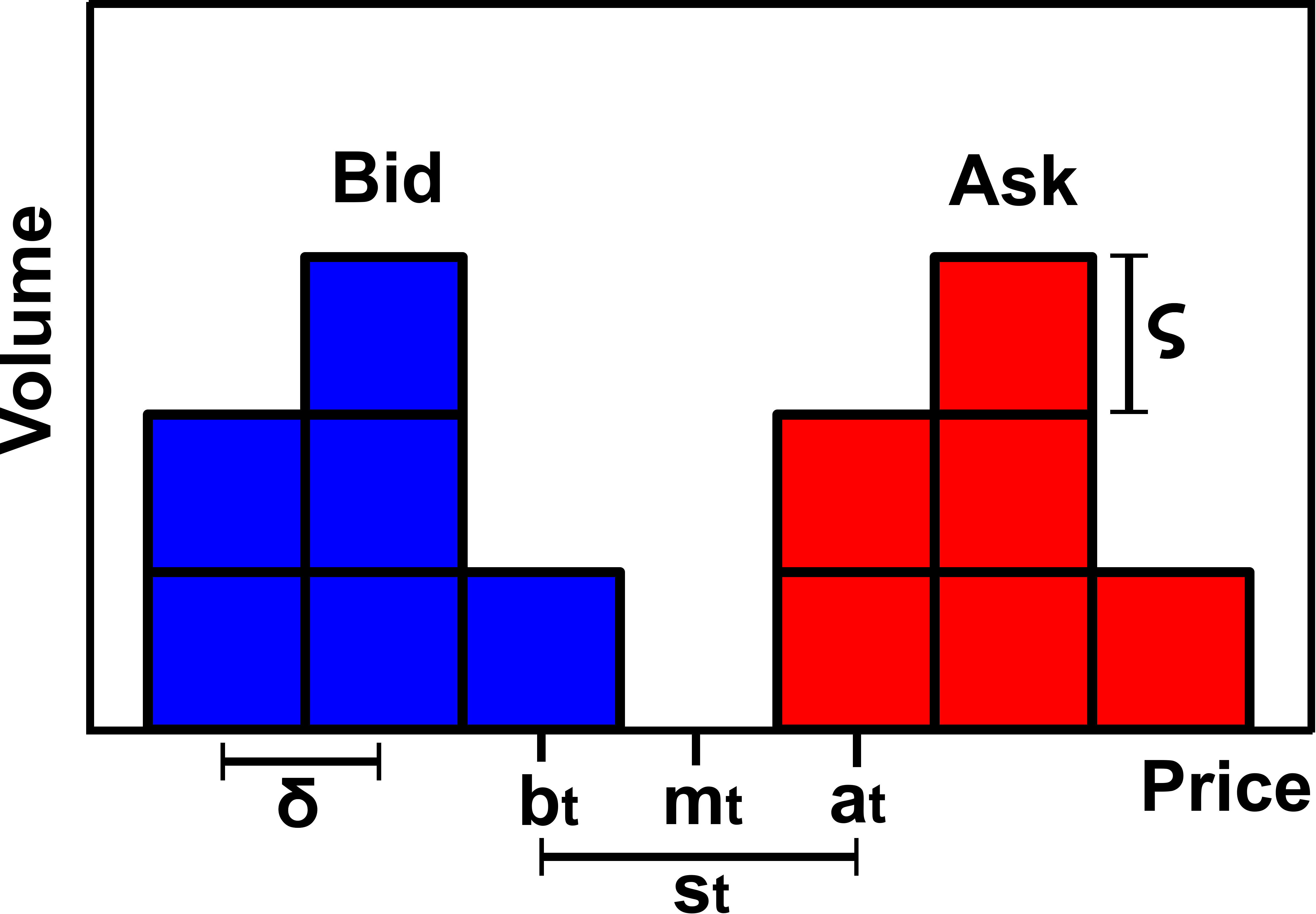}
\caption{\textbf{Schematic of a LOB and related terminology.} At any time $t$ the \emph{bid price} $b_t$ is the highest limit price among all the buy limit orders (blue) while the \emph{ask price} $a_t$ is the lowest limit price among all the sell limit orders (red). The bid and ask prices are the \emph{best quotes} of the LOB. The mid point between the best quotes $m_t = (a_t + b_t)/2$ is the \emph{mid price}. The distance between the best quotes $s_t = a_t - b_t$ is the \emph{bid-ask spread}. The volume specified in a limit order must be a multiple of the \emph{lot size} $\varsigma$, which is the minimum exchangeable quantity (in units of the traded asset). The price specified in a limit order must be a multiple of the \emph{tick size} $\delta$, which is the minimum price variation imposed 
by the LOB. The lot size $\varsigma$ and the tick size $\delta$ are known as \emph{resolution parameters} of the LOB \cite{gould2013limit}. Orders are allocated in the LOB depending on their distance (in multiples of $\delta$) from the current best quote. For instance, a buy limit order with price $b_t - n\delta$ occupies the $n + 1$-th level of the bid side.}
\end{figure}
The limit order with the best price (i.e., the highest bid or the lowest ask quote) is always the first to be matched against a forthcoming order. The adoption of a minimum price increment $\delta$ forces the price to move in a discrete grid, hence the same price can be occupied by multiple limit orders at the same time. As a result, exchanges adopt an additional rule to prioritize the execution of orders bearing the same price. A very common scheme is the \emph{price-time} priority rule which uses the submission time to set the priority among limit orders occupying the same price level, i.e. the order that entered the LOB earlier is executed first \cite{bonart2018continuous}. 
\subsubsection{Triangular Arbitrage}\label{sec:triangularbitrage}
\begin{figure}[H]
\centering
\includegraphics[scale=0.125]{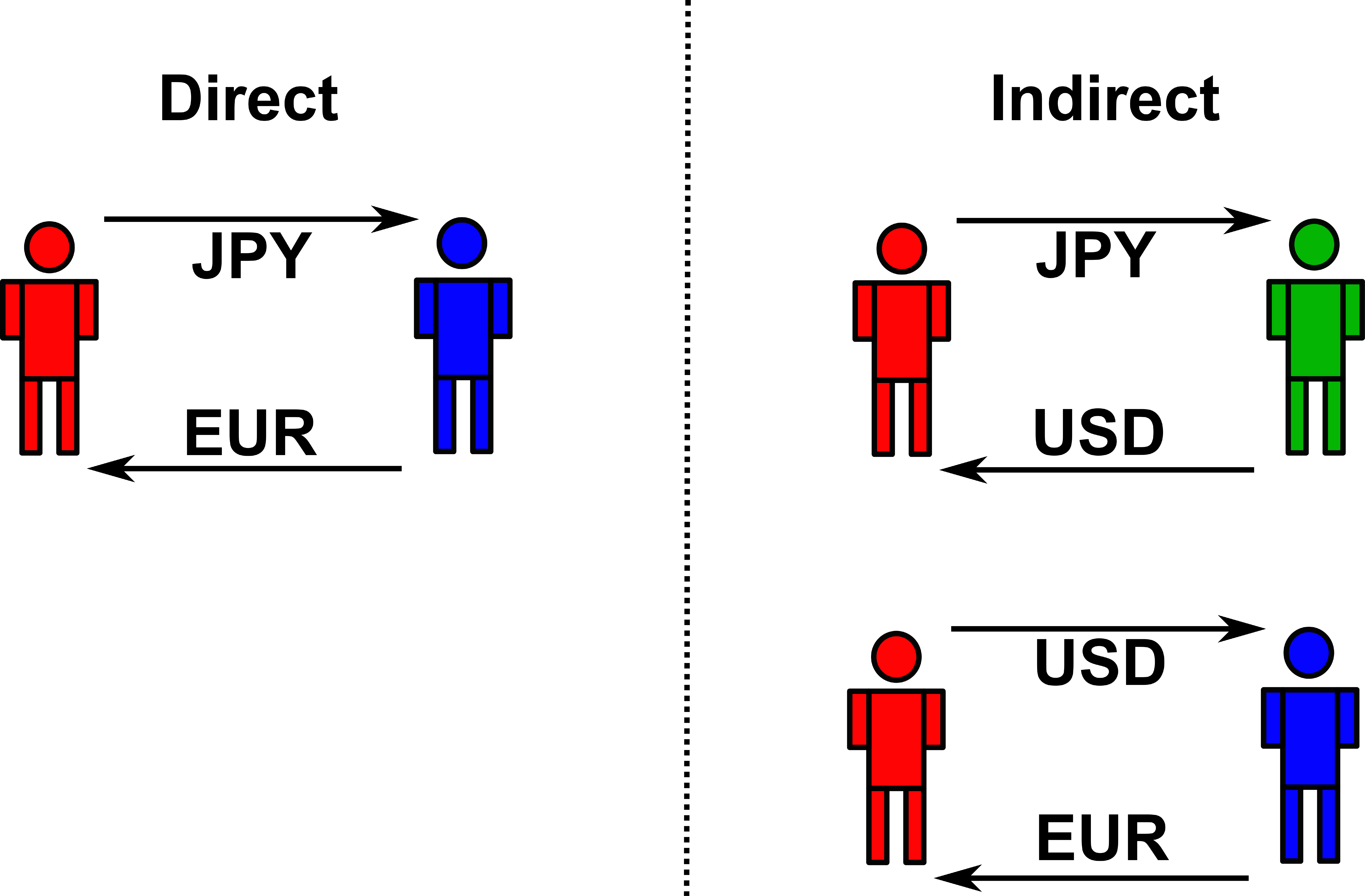}
\caption{\textbf{Two ways of obtaining one unit of EUR.} Direct transaction (left panel): agent \#1 (red) obtains EUR from agent \#2 (blue) in exchange for JPY. Indirect transaction (right panel): agent \#1 purchases USD from agent \#3 (green) in exchange for JPY. Then, she obtains EUR from agent \#2 (blue) in exchange for USD.}
\label{fig:scheme}
\end{figure}

In the FX market, the price of a currency is always expressed in units of another currency and it is commonly known as foreign exchange rate (FX rate henceforth). For instance, the price of one Euro (EUR henceforth) in Japanese Yen (JPY henceforth) is denoted by EUR/JPY. The same FX rate can be obtained from the product of two other FX rates, e.g. EUR/JPY = USD/JPY$\times$EUR/USD, where USD indicates US Dollars. In the former case EUR is purchased directly while in the latter case EUR is purchased indirectly through a third currency (i.e., USD), see Fig.~\ref{fig:scheme}.
At any time $t$ we expect the following equality to hold
\begin{equation}\label{parity}
    \underbrace{\text{EUR}/\text{JPY}_{t}}_{\text{FX rate}} = \underbrace{(\text{USD}/\text{JPY}_{t})\times(\text{EUR}/\text{USD}_{t})}_{\text{implied FX cross rate}},
\end{equation}
that is, the costs of a direct and indirect purchase of the same amount of a given currency must be the same. Clearly, Eq.~\eqref{parity} can be generalized to any currency triplet. \\
However, several datasets \cite{aiba2002triangular,marshall2008exploitable,fenn2009mirage,foucault2016toxic} reveal narrow time windows in which Eq.~\eqref{parity} does not hold. In this scenario, traders might try to exploit one of the following misprices
\begin{subequations}
\begin{equation}\label{missmatch1}
    \text{EUR}/\text{JPY}_{t} < (\text{USD}/\text{JPY}_{t})\times(\text{EUR}/\text{USD}_{t})
\end{equation}
\begin{equation}\label{missmatch2}
    \text{EUR}/\text{JPY}_{t} > (\text{USD}/\text{JPY}_{t})\times(\text{EUR}/\text{USD}_{t})
\end{equation}
\end{subequations}
by implementing a triangular arbitrage strategy. For instance, Eq.~\eqref{missmatch2} suggests that a trader holding JPY could gain a risk-free profit by buying EUR indirectly (JPY $\rightarrow$ USD $\rightarrow$ EUR) and selling EUR directly (EUR $\rightarrow$ JPY). 
 \begin{figure}[H]
\centering
\includegraphics[scale = 0.10]{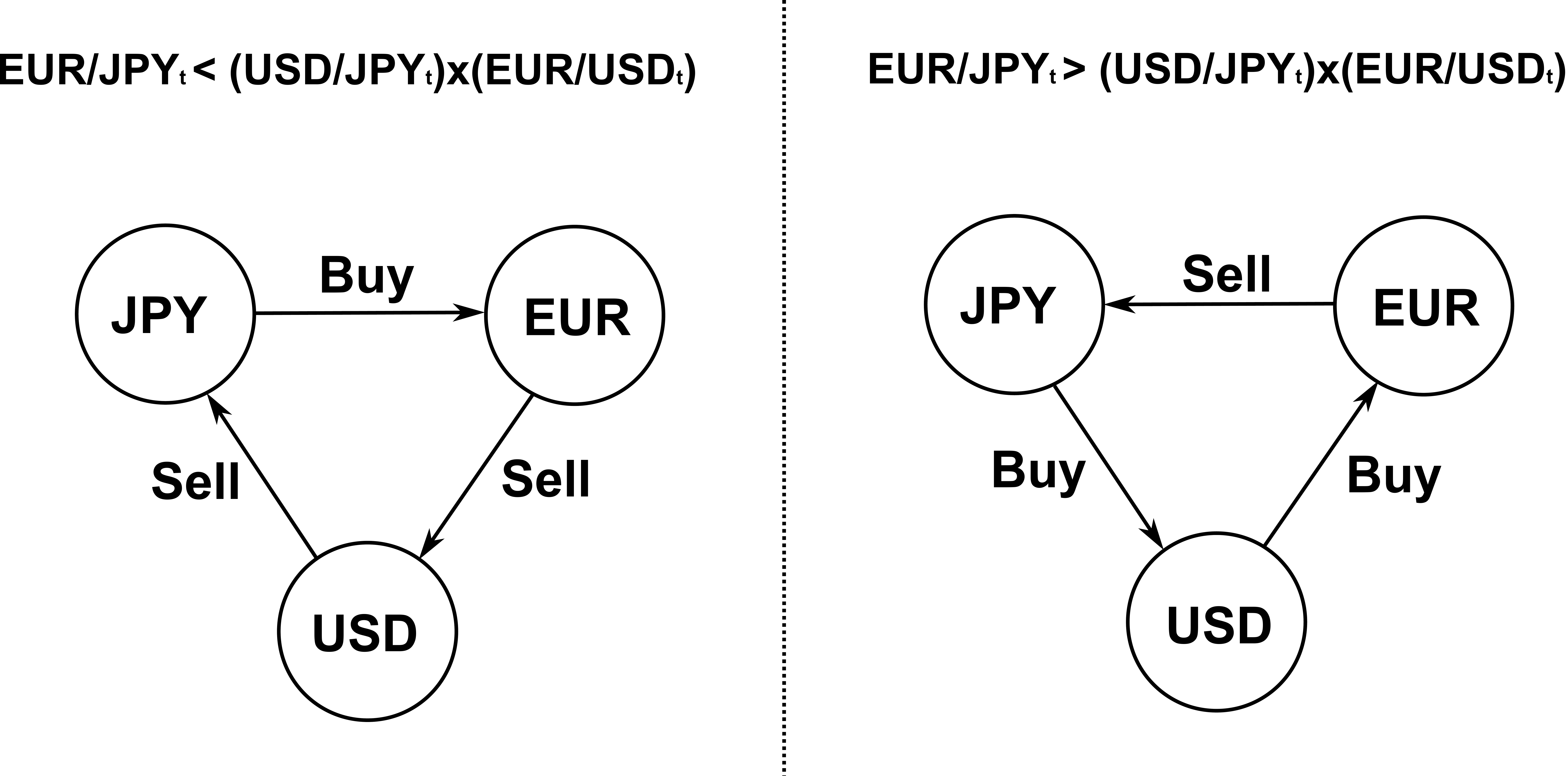}
\caption{\textbf{Profitable misprices and associated triangular arbitrage strategies}. Left panel: an agent buys EUR for JPY and sells EUR for JPY through USD. Right panel: an agent sells EUR for JPY and buys EUR for JPY through USD.}
\label{fig:triarb}
\end{figure}
Assuming that the arbitrager completes each transaction at the best quotes (i.e., sell at the best bid and buy at the best ask) available in the EUR/JPY, USD/JPY and EUR/USD LOBs, any strategy presented in Fig.~\ref{fig:triarb} is effectively profitable if the following condition (i.e., Eq.~\eqref{micromissmatch1} for left panel strategy or Eq.~\eqref{micromissmatch2} for right panel strategy) is satisfied
\begin{subequations}
        \begin{equation}\label{micromissmatch1}
            a_{ \text{EUR}/\text{JPY}}(t) < b_{\text{USD}/\text{JPY}}(t)\times b_{\text{EUR}/\text{USD}}(t)
        \end{equation}
        \begin{equation}\label{micromissmatch2}
            b_{ \text{EUR}/\text{JPY}}(t) > a_{\text{USD}/\text{JPY}}(t)\times a_{\text{EUR}/\text{USD}}(t)
        \end{equation}
        \end{subequations}
where $b_{x/y}(t)$ and $a_{x/y}(t)$ are the best bid and ask quotes available at time $t$ in the $x/y$ market respectively.\\
In the same spirit of \cite{aiba2002triangular,aiba2006microscopic,fenn2009mirage}, we detect the presence of triangular arbitrage opportunities when one of the following processes
\begin{subequations}
            \begin{equation}\label{mu1}
            \mu^{I}(t) = \frac{b_{\text{USD}/\text{JPY}}(t)\times b_{\text{EUR}/\text{USD}}(t)}{a_{\text{EUR}/\text{JPY}}(t)}
            \end{equation}
            \begin{equation}\label{mu2}
            \mu^{II}(t) = \frac{b_{\text{EUR}/\text{JPY}}(t)}{a_{\text{USD}/\text{JPY}}(t)\times a_{\text{EUR}/\text{USD}}(t)}
            \end{equation}
\end{subequations}
exceeds the unit.
\subsection{The EBS Dataset}\label{sec:data}
In this study, we employ highly granular LOB data provided by Electronic Broking Services (EBS henceforth). EBS is an important inter-dealer electronic platform for FX spot trading \cite{BIS}. Trading is organized in LOBs and estimates suggest that approximately 70\% of the orders are posted by algorithm \cite{BIS}. 
We investigate three major FX rates, USD/JPY, EUR/USD and EUR/JPY, across four years of trading activity (2011-2014).  
\begin{table}[H]
\caption{\textbf{EBS dataset structure.}}
 \centering
 \resizebox{\columnwidth}{!}{%
 \begin{tabular}{|c|c|c|c|c|c|c|c|}
   \hline
   \shortstack{Date} & \shortstack{Timestamp} & \shortstack{Market} & \shortstack{Event} & \shortstack{Direction} & \shortstack{Depth} & \shortstack{Price} & \shortstack{Volume}\\
   \hline
   2011-05-10 & 09.00.00.000 & USD/JPY & Deal & Buy & 1st & 100.000 & 1\\
   \vdots&\vdots & \vdots & \vdots & \vdots & \vdots & \vdots & \vdots \\
   2011-10-21 & 21.00.00.000 & EUR/USD & Quote & Ask & 3rd & 0.8000 & 5 \\
   \hline
 \end{tabular}%
 }
  \caption*{
  \\
  Each record (i.e., row) corresponds to a specific market event. Records are reported in chronological order (top to bottom) and include the following details: i) date (yyyy-mm-dd), ii) timestamp (GMT), iii) the market in which the event took place, iv) event type (submission (Quote) or execution (Deal) of visible or hidden limit orders), v) direction of limit orders (Buy/Sell for deals and Bid/Ask for quotes), vi) depth (number of occupied levels) between the specified price and the best price, vii) price and viii) units specified in the limit order. }
\label{tab:EBSScheme}
\end{table}
The shortest time window between consecutive records is 100 millisecond (ms). 
Events occurring within 100 ms are 
aggregated and recorded at the nearest available timestamp. The tick size has changed two times within the considered four years window, see \cite{mahmoodzadeh2014tick} and Table \ref{tab:ticksizes} for further details.\\
The EBS dataset provides a 24-7 coverage of the trading activity 
(from 00:00:00.000 GMT Monday to 23:59:59.999 GMT Sunday included), 
thus offering a complete and uninterrupted record 
of the flow of submissions, executions and deletions occurring in the first ten price levels of the bid and ask sides of the LOB.
\\
The EBS dataset, in virtue of its features, is a reliable source of granular market data. 
First, EBS directly collects data from its own trading platform. This prevents the common issues associated to the presence of third parties during the recording process, such as interpolations of missing data and input errors (e.g., incorrect timestamps or order types).
Second, the EBS dataset offers a continuous record of LOB events across a wide spectrum of currencies, thus becoming a natural choice for cross-sectional studies (e.g., triangular arbitrage or correlation networks). Third, in spite of the increasing competition, the EBS platform has remained a key channel for accessing FX markets for more than two decades by connecting traders across more than 50 countries \cite{EBShome,EBSplat}. The enduring relevance of this platform has been guaranteed by the fairness and the competitiveness of the quoted prices. \\
\subsection{The Arbitrager Model}\label{sec:Model}
We introduce a new microscopic model (Arbitrager Model hencefort) in which market makers trade $d = 3$ FX rates in $d = 3$ inter-dealer markets. Trading is organized in LOBs and, for simplicity, prices move in a continuous grid. We enforce the assumption that market makers cannot interact across markets, that is, they can only trade in the LOB they have been assigned to. Finally, echoing \cite{aiba2006microscopic}, we include a special agent (i.e., the arbitrager) that is allowed to submit market orders in any market.
\begin{figure}[H]
	\centering
	\includegraphics[scale=0.05]{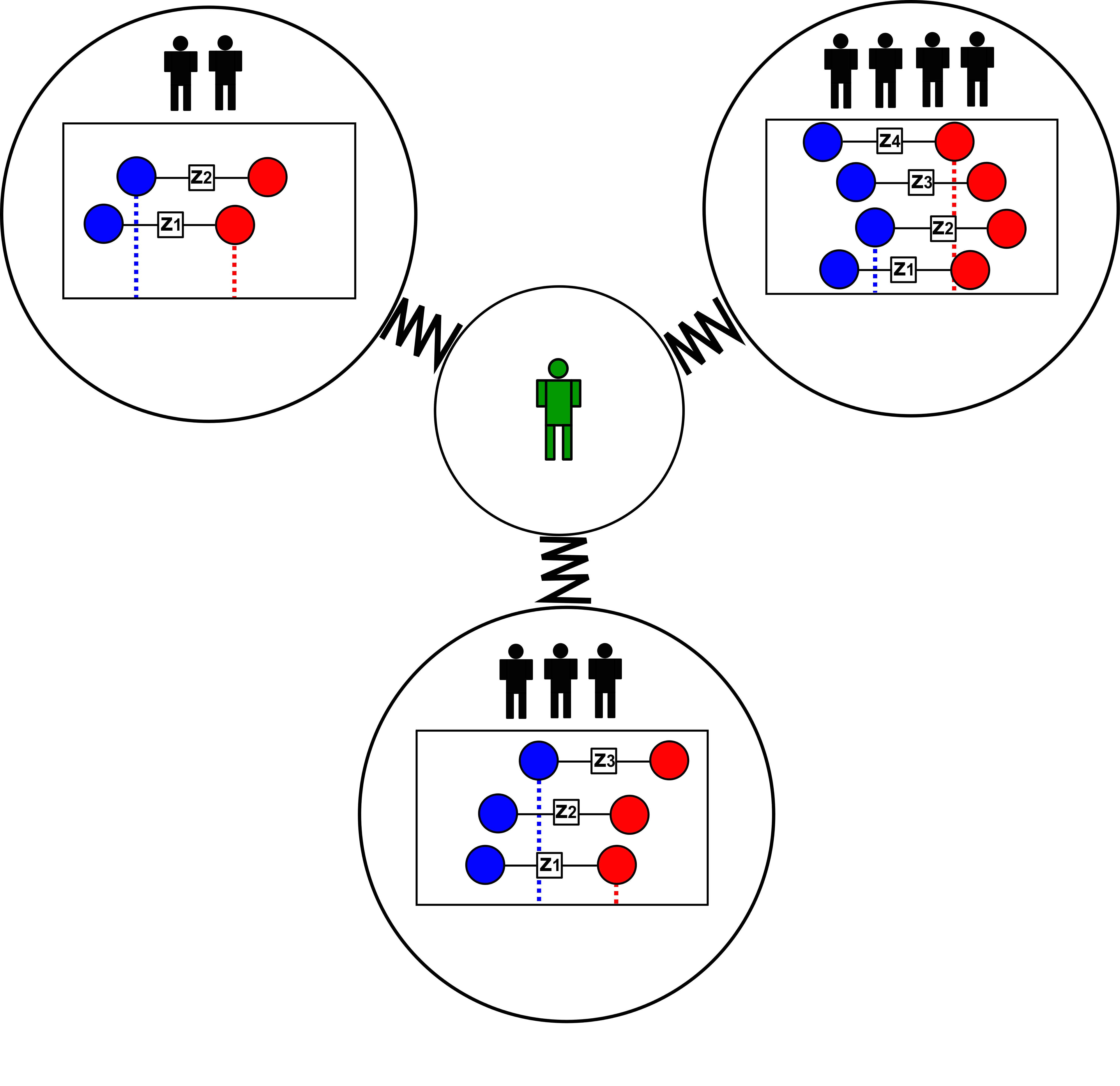}
	\caption{\textbf{Schematic of the Arbitrager Model ecology.} The ecology comprises three independent markets. Trading is organized in continuous price grid LOBs as in \cite{yamada2009solvable}, see Section \ref{appendix:yamadamodel}. Market makers (black agents) maintain bid (blue circles) and ask (red circles) quotes with constant spread (black segment). To adjust these quotes, market makers dynamically update their dealing prices (squares) by adopting trend-based strategies. The best quotes are marked by dotted lines (blue for bid and red for ask). Transactions occur when the best bid matches or exceeds the best ask. Market makers engaging in a trade close the deal at the mid point between the two matching prices (i.e., transaction price), see Fig.~\ref{arbitragermodelinteractions}. Finally, we include an arbitrager (green agent) that exclusively submits market orders across the three markets to exploit triangular arbitrage opportunities emerging now and then. The actions of this special agent, affecting the events occurring in otherwise independent markets, entangle the dynamics of the FX rates traded in the ecology. Echoing \cite{aiba2006microscopic}, the ecology can be visualized as a spring-mass system in which the dynamics of three random walkers (i.e., the markets) are constrained by a restoring force (i.e., the arbitrager) acting on the center of gravity of the system.
	}
	\label{fig:DMEcology}
\end{figure}
\subsubsection{Market makers}\label{model:marketmakers}
The $i$-th market maker operating in the $\ell$-th market actively manages a bid quote $b_{i,\ell}(t)$ and an ask quote $a_{i,\ell}(t)$ separated by a constant spread $L_{\ell} = a_{i,\ell}(t) - b_{i,\ell}(t)$. To do so, the $i$-th market maker updates its dealing price $z_{i,\ell}(t)$ , which is the mid point between the two quotes (i.e.,  $z_{i,\ell}(t) = a_{i,\ell}(t) - L_{\ell}/2 = b_{i,\ell}(t) + L_{\ell}/2$), by adopting a trend-based strategy
\begin{equation}\label{midpriceeq2}
z_{i,\ell}(t) = z_{i,\ell}(t - dt) + c_{\ell}\phi_{n,\ell}(t)dt + \sigma_{\ell}\sqrt{dt}\epsilon_{i,\ell}(t) \;\;\; i = 1,\dots, N_{\ell}
\end{equation}
where $N_{\ell}$ is the number of market makers participating the $\ell$-th market, $\sigma_{\ell} > 0$, and $\epsilon_{i,\ell}(t) \sim \mathcal{N}(0,1)$. The term
\begin{equation}\label{trend2}
\phi_{n,\ell}(t)= \frac{\sum\limits_{k = 0}\limits^{n-1}\left(p_{\ell}(g_{t, \ell} - k)-p_{\ell}(g_{t,\ell} -k - 1)\right)e^{-\frac{k}{\xi}}}{\sum\limits_{k = 0}\limits^{n-1}e^{-\frac{k}{\xi}}} \;\;\; \ell = 1,\dots, d
\end{equation}
is the weighted average of the last $n < g_{t,\ell}$ changes in the transaction price $p_{\ell}$ in the $\ell$-th market, $g_{t,\ell}$ is the number of transactions occurred in $[0,t[$ in the $\ell$-th market and $\xi > 0$ is a constant term. The real-valued parameter $c_{\ell}$ controls how the current price trend $\phi_{n,\ell}(t)$ influences market makers' strategies. For instance, $c_{\ell} > 0$ ($c_{\ell} < 0$) indicates that market makers operating in the $\ell$-th market tend to adjust their dealing prices $z(t)$ in the same (opposite) direction of the sign of the price trend $\phi_{n,\ell}(t)$.\\
Transactions occur when the $i$-th market maker is willing to buy at a price that matches or exceeds the ask price of the $j$-th market maker (i.e., $b_{i,\ell} \geq a_{j,\ell}$). Trades are settled at the transaction price $p(g_{t,\ell}) = (a_{j,\ell}(t) + b_{i,\ell}(t))/2$ and only the market makers who have just engaged in a trade adjust their dealing prices $z(t + dt)$ to the latest transaction price $p(g_{t,\ell})$, see Figs.~\ref{dealermodelinteractions} and \ref{arbitragermodelinteractions} 
\subsubsection{The arbitrager}\label{model:arbitrager}
The arbitrager is a liquidity taker (i.e., she does not provide bid and ask quotes like market makers) that can only submit market orders in each market to exploit an existing triangular arbitrage opportunity. Assuming that agents exchange EUR/JPY, EUR/USD and USD/JPY, the triangular arbitrage processes are
\begin{subequations}
\begin{equation}\label{dmarb1}
\mu^{I}(t) = \frac{b_{\text{EUR/USD}}(t)\times b_{\text{USD/JPY}}(t)}{a_{\text{EUR/JPY}}(t)} 
\end{equation}
\begin{equation}\label{dmarb2}
\mu^{II}(t) = \frac{b_{\text{EUR/JPY}}(t)}{a_{\text{EUR/USD}}(t)\times a_{\text{USD/JPY}}(t)}
\end{equation}
\end{subequations}
where $b_{\ell}(t)$ and $a_{\ell}(t)$ are the best bid and ask quotes at time $t$ in the $\ell$-th market.
\\
Whenever Eqs.~\eqref{dmarb1} or \eqref{dmarb2} exceeds the unit, the arbitrager submits market orders to exploit the current opportunity (henceforth predatory market orders). Contrary to limit orders, market orders trigger an immediate transaction between the arbitrager and the market maker providing the best quote on the opposite side of the LOB. This implies that transactions involving the arbitrager are always settled at the bid or ask quote offered by the matched market maker, which are by the definition the current best bid or ask quote of the LOB. Following the post-transaction update rule, the matched market maker adjust its dealing price to its own matched bid or ask quote, that is, $z_{i,\ell}(t + dt) \rightarrow a_{i,\ell}(t)$ in case of a buy predatory market order or $z_{i,\ell}(t + dt) \rightarrow b_{i,\ell}(t)$ in case of a sell predatory market order, see Fig.~\ref{fig:arbitragermodelattack}.
\section{Results}\label{sec:results}
\subsection{Cross-correlation functions}\label{sec:results1}
Echoing previous empirical studies \cite{mizuno2004time,wang2013statistical}, we examine the shape of the cross-correlation function
\begin{subequations}
\begin{equation}\label{crosscorr}
\rho_{i,j}(\omega) = \frac{\langle \Delta m_{i}(t)\Delta m_{j}(t)\rangle - \langle\Delta m_{i}(t)\rangle\langle\Delta m_{j}(t)\rangle}{\sigma_{\Delta m_{i}}\sigma_{\Delta m_{j}}}
\end{equation}
\begin{equation}
    \sigma_{\Delta m_{\ell}} = \left(\langle\Delta m_{\ell}(t)^2 \rangle - \langle\Delta m_{\ell}(t) \rangle^2\right)^{1/2}
\end{equation}
\end{subequations}
where the time-scale $\omega$ is the interval (i.e., in seconds) between two consecutive observations of the $\ell$-th mid price $m_{\ell}$ time series, 
$\Delta m_{\ell}(t) \equiv m_{\ell}(t) - m_{\ell}(t - \omega)$ is the linear change between consecutive observations and $\sigma_{\Delta m_{\ell}}$ is the standard deviation of $\Delta m_{\ell}(t)$. 
\begin{figure}[H]
  \begin{center}
    \includegraphics[width=13cm]{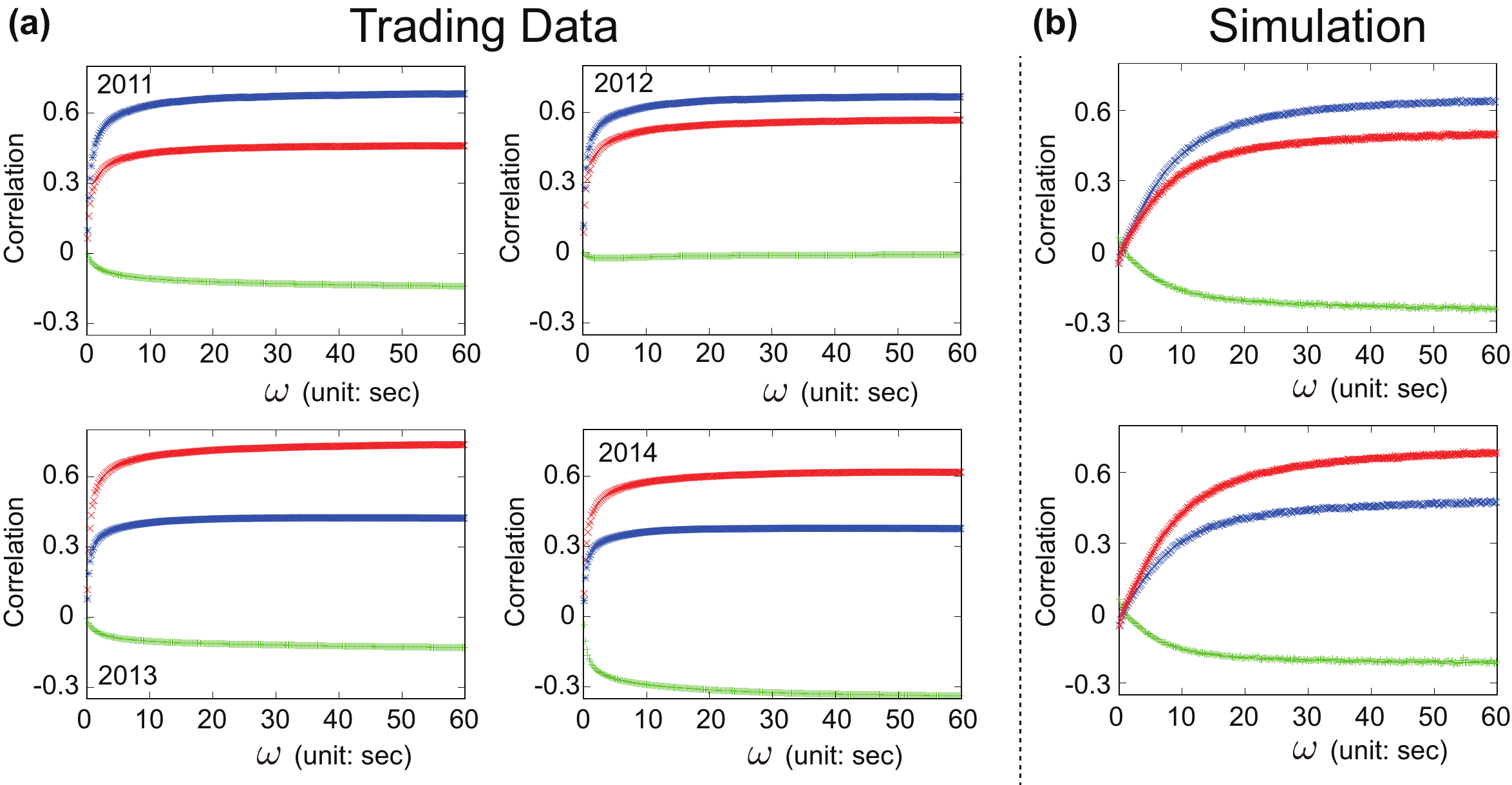}
    \caption{\textbf{Trading data vs. model based cross-correlation functions.} Cross-correlation function $\rho_{i,j}(\omega)$ for $\Delta$USD/JPY vs. $\Delta$EUR/USD (green), $\Delta$EUR/USD vs. $\Delta$EUR/JPY (blue) and $\Delta$USD/JPY vs. $\Delta$EUR/JPY (red) as a function of the time-scale $\omega$ of the underlying time series. (a) Real market data (EBS) across four distinct years (2011-2014). (b) Arbitrager Model simulations. The number of participating market makers $(N_{\text{EUR/USD}},N_{\text{USD/JPY}},N_{\text{EUR/JPY}})$ are $(35,45,25)$ in the first experiment, see (b) top panel, and $(50,35,25)$ in the second experiment, see (b) bottom panel. The trend-following strength parameters are $(c_{\text{EUR/USD}},c_{\text{USD/JPY}},c_{\text{EUR/JPY}}) = (0.8, 0.8, 0.8)$ in both experiments. The length of each simulation is $5\times 10^6$ time steps. The price trends $\phi_{n,\ell}$ are calculated over the most recent $n = 15$ changes in the transaction price $p$ and the scaling constant is set to $\xi = 5$, see Eq.~\eqref{trend2}. Details on the initialization of the model and the conversion between simulation time (i.e., time steps) and real time (i.e., sec) are provided in Section \ref{appendix:calibration}.} 

    \label{fig:TwoCorr}
  \end{center}
\end{figure}
In real trading data we observe that the value of the cross-correlation function $\rho_{i,j}(\omega)$ varies with $\omega$ on very short time-scales ($\omega < 1$ sec). This time-scale dependency starts to weaken after $\omega \approx 1$ sec and vanishes beyond $\omega \approx 10$ sec, see Fig.~\ref{fig:TwoCorr}(a). 
The characteristic shape of $\rho_{i,j}(\omega)$ displayed in Fig.~\ref{fig:TwoCorr}(a) is compatible with the one found by Mizuno \textit{et al.} \cite{mizuno2004time}. However, our trading data-based cross-correlation functions stabilize on much shorter time-scales. Considering that \cite{mizuno2004time} employed trading data collected in 1999, a period where lower levels of automation imposed a slower trading pace, we hypothesize that the time-scale $\omega$ beyond which $\rho_{i,j}(\omega)$ stabilizes reflects the speed at which markets react to a given event. Furthermore, $\rho_{i,j}(\omega)$ stabilizes around different levels over the four trading years covered in our analysis. For instance, the cross-correlation between $\Delta$USD/JPY and $\Delta$EUR/JPY, see Fig.~\ref{fig:TwoCorr}(a), stabilizes around 0.6 in 2011-2012 and 0.3 in 2013-2014. We assert that the variability in the stabilization levels of $\rho_{i,j}(\omega)$ might be related to the different tick sizes adopted by EBS during the four years covered in this empirical analysis, see \cite{mahmoodzadeh2014tick} and Table~\ref{tab:ticksizes}. Detailed investigations on how changes in the design of FX LOBs (e.g., tick size) and the increasing sophistication of market participants (e.g., high frequency traders) affect the characteristic shape of $\rho_{i,j}(\omega)$ are outside the scope of this paper, however, such studies will be a very much welcomed addition to the current literature.\\
The Arbitrager Model satisfactorily replicates the characteristic shape of $\rho_{i,j}(\omega)$, suggesting that triangular arbitrage plays a primary role in the entanglement of the dynamics of currency pairs in real FX markets.
We observe two quantitative differences between the characteristic shape of $\rho_{i,j}(\omega)$ derived from simulations of the Arbitrager Model and real trading data. First, $\rho_{i,j}(\omega)$ flattens after $\omega \approx 30$ sec in the model, see Fig.~\ref{fig:TwoCorr}(b), and $\omega \approx 10$ sec in real trading data, see Fig.~\ref{fig:TwoCorr}(a). Second, in extremely short time-scales ($\omega \rightarrow 0$ sec) the model-based $\rho_{i,j}(\omega)$ does not converge to zero as in real trading data, see Fig.~\ref{fig:TwoCorr}(b), but to nearby values. We assert that these discrepancies stem from the extreme simplicity of the Arbitrager Model which neglects various practices of real FX markets that contribute, to different degrees, to the shape and features of $\rho_{i,j}(\omega)$ revealed in real trading data. To support this hypothesis, we developed an extended version of the Arbitrager Model which includes additional features of real FX markets, see Section \ref{appendix:extended}. This more complex version of our model overcomes the main differences between the curves displayed in Fig.~\ref{fig:TwoCorr}(a) and (b), reproducing cross-correlation functions $\rho_{i,j}(\omega)$ that approach zero when $\omega \rightarrow 0$ sec and stabilize on shorter time-scales than those emerged in the baseline model.
\subsection{The interplay between triangular arbitrage and trend-following strategies intertwines FX rates dynamics}\label{sec:results2}
\begin{figure}[H]
  \begin{center}
    \includegraphics[width=13cm]{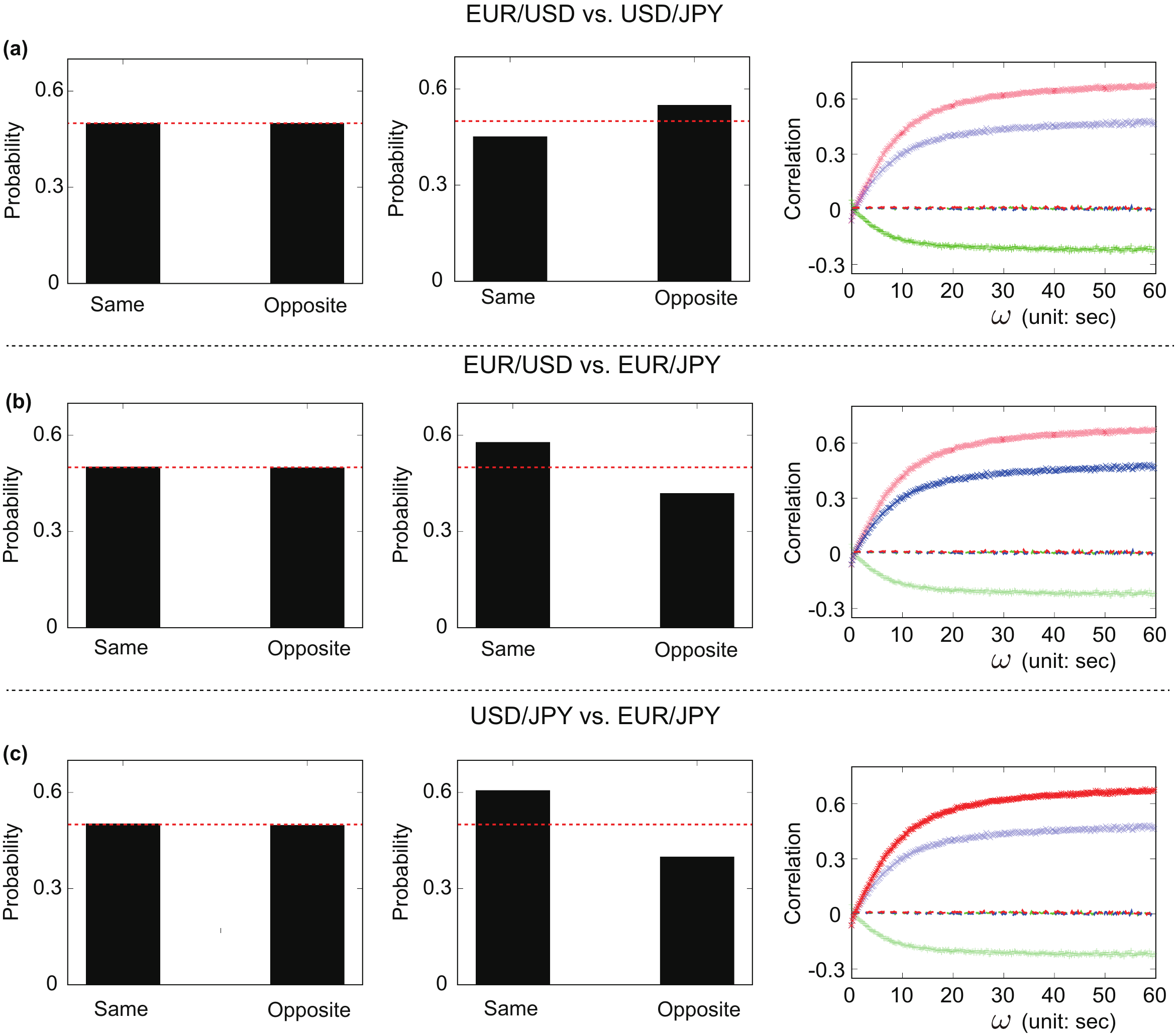}
    \caption{\textbf{Statistical relationships between different FX markets.} Probability of observing two markets in the same or opposite state in the absence of the arbitrager (left column), with the arbitrager (central column) and the associated cross-correlation functions $\rho_{i,j}(\omega)$ (right column) for (a) $\Delta$EUR/USD vs. $\Delta$USD/JPY, (b) $\Delta$EUR/USD vs. $\Delta$EUR/JPY and (c) $\Delta$USD/JPY vs. $\Delta$EUR/JPY. The red solid line in the histograms marks the value of 0.5, highlighting the case in which two markets have the same probability of being in the same or opposite state. The lines indicating the value of the cross-correlation function $\rho_{i,j}(\omega)$ are solid (dashed) for experiments including (excluding) the arbitrager. Simulations are performed under the same settings of the experiment presented in Fig.~\ref{fig:TwoCorr}(b), bottom panel. We find that the inclusion of the arbitrager increases the probability of observing EUR/USD and USD/JPY as well as EUR/USD and EUR/JPY in the same state and USD/JPY and EUR/USD in the opposite state. Furthermore, the active presence of this special agent intertwines the dynamics of different FX rates, creating cross-correlations functions that resemble those emerged in real trading data.} 
    \label{fig:Heterogenuity}
  \end{center}
\end{figure}

\begin{figure}[H]
	\centering
	\includegraphics[scale=0.75]{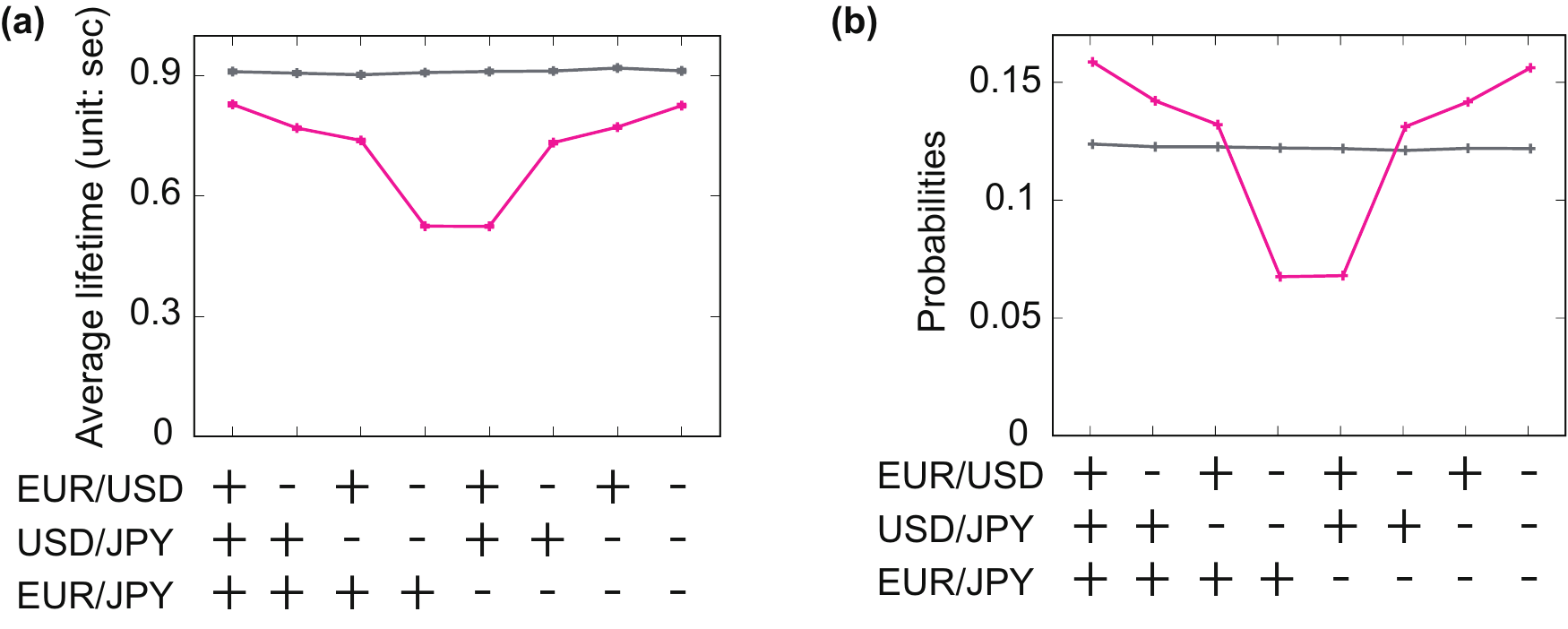}
	\caption{\textbf{Expected lifetime and appearance probability of the eight ecology configurations.} Statistics are collected from simulations of the Arbitrager Model with active (violet) and inactive (grey) arbitrager. Simulations are performed under the same settings of the experiment presented in Fig.~\ref{fig:TwoCorr}(b), bottom panel. We observe that the presence of an active arbitrager increases the average lifetimes (a) and appearance probabilities (b) of certain configurations and reduces the same statistics for others. Statistics in (a) are expressed in real time (i.e., sec.), details on the conversion between simulation time (i.e., time steps) and real time (i.e., sec) are provided in Section \ref{appendix:calibration}.
	}
	\label{fig:statusDuration}
\end{figure}

The Arbitrager Model, reproducing the characteristic shape of $\rho_{i,j}(\omega)$, suggests that triangular arbitrage plays a primary role in the formation of the cross-correlations among currencies. However, it is not clear how the features of $\rho_{i,j}(\omega)$, such as its sign and values, stem from the interplay between the different types of strategies adopted by agents operating in our ecology. Addressing this open question is one of the main objectives of the present study.\\
We define the actual state of the $j$-th market $\nu_{j}(t)$ as the sign of the current price trend $\text{sgn}(\phi_{n,\ell}(t)) \in \{-, +\}$, see Eq.~\eqref{trend2}. It follows that the current configuration of the ecology $q(t) = \{\nu_{1}(t), \nu_{2}(t), \nu_{3}(t)\}$ is the combination of the states of each market. Our model, considering three markets, admits $2^3 = 8$ different ecology configurations.
When the arbitrager is not included in the system, two markets have the same probability of being in the same and opposite state, see first column of Fig.~\ref{fig:Heterogenuity}. This occurs because price trends are driven
by transactions triggered by endogenous decisions, that is, events occurring in different markets remain completely unrelated. As a consequence, market states flip independently and at the same rate. It follows that the eight possible combinations of market states share the same appearance probabilities $1/2^3$ and expected lifetimes, see Fig.~\ref{fig:statusDuration}. In these settings, the dynamics of the mid price of FX rate pairs do not present any significant correlation, see third column of Fig.~\ref{fig:Heterogenuity}.
The inclusion of the arbitrager has a major impact on the overall behavior of the model. We notice the emergence of imbalances in the probability of observing two markets in the same or opposite state. For instance, the EUR/USD and EUR/JPY markets have the same state in $\approx$ 57\% of the experiment duration, see Fig.~\ref{fig:Heterogenuity}(b). Movements of FX rate pairs become correlated, revealing cross-correlation functions $\rho_{i,j}(\omega)$ whose shapes qualitatively mimic those found in real trading data. The sign and stabilization levels of these functions are consistent with the sign and size of the probabilities imbalances, suggesting that these two results are two faces of the same coin.\\
To understand how the findings presented in Fig.~\ref{fig:Heterogenuity} unfold we need to take a closer look at the statistical properties of the eight ecology configurations.
The presence of the arbitrager introduces a degree of heterogeneity in both the expected lifetimes and appearance probabilities of ecology configurations, see Fig.~\ref{fig:statusDuration}. This reveals three interesting facts. First, the average lifetime of every ecology configuration is smaller than its counterpart in an arbitrager-free system. To explain this feature, we recall that predatory market orders trigger three simultaneous transactions (i.e., one in each market) altering the current price trends $\phi_{n,\ell}(t)$, see Eq.~\eqref{trend2}. When the latest change in transaction price $p_{\ell}(g_{t,\ell}) - p_{\ell}(g_{t,\ell} - 1)$ induced by a predatory market order and $\phi_{n,\ell}(t - dt)$ have opposite signs, the actions of the arbitrager weaken (i.e., $|\phi_{n,\ell}(t)| < |\phi_{n,\ell}(t - dt)|$) or even flip the sign (i.e., $\phi_{n,\ell}(t)\phi_{n,\ell}(t - dt) < 0$) of the price trend. When this occurs, the arbitrager weakens the trend-following behaviors of market makers in at least one of the three markets, thus increasing the likelihood of a transition to another ecology configuration. As triangular arbitrage opportunities of both types appear, with different incidences, during any ecology configuration, see Fig.~\ref{fig:mubarplot}, the expected lifetimes of these configurations are, to different extents, shorter than in an arbitrager-free system.
Second, certain ecology configurations are expected to last more than others (i.e., single episodes). As reactions to triangular arbitrage opportunities increase the likelihood of flipping a market state, the average lifetime of a given configuration relate to the time required for the first triangular arbitrage opportunity to emerge. For instance, the time between the inception and the first time $\mu^{I}(t)$ or $\mu^{II}(t)$ becomes larger than one never exceeds 4 sec for $\{-,-,+\}$, which is the configuration with shortest expected lifetime, while it can reach $\approx$ 6.5 sec for $\{+,+,+\}$, which is the configuration with longest expected lifetime, see Fig.~\ref{fig:prearbitrageccdf}(a). This difference can be intuitively explained by looking at the combination of market states. When the ecology configuration is $\{-,-,+\}$, EUR/USD and USD/JPY have the opposite state of EUR/JPY. In this scenario, the implied FX cross rate EUR/USD$\times$USD/JPY moves in the opposite direction of the FX rate EUR/JPY, creating the ideal conditions for a rapid emergence of triangular arbitrage opportunities. Conversely, the three markets share the same state when the ecology configuration is $\{+,+,+\}$. In this case, both the FX rate and the implied FX cross rate move in the same direction, extending the time required by these prices to create a gap that can be exploited by the arbitrager. 
\\
The third and final interesting fact emerged in Fig.~\ref{fig:statusDuration} is that certain configurations are more likely to appear than others. To understand this aspect, we first highlight the significant differences between the probabilities of transitioning from a configuration to another, see Table~\ref{tab:tranmatrix}. For instance, assuming that the system is leaving $\{+,+,+\}$, the probabilities of transitioning to $\{-,+,+\}$ and $\{+,+,-\}$ are 35.8\% and 22.7\%, respectively. This difference can be explained by the fact that it is much easier to flip the state of EUR/USD and move to $\{-,+,+\}$ than flipping EUR/JPY and move to $\{+,+,-\}$. The value of the price trend $\phi_{n,\ell}(t)$ can be intuitively seen as the \emph{resistance} to state changes of the $\ell$-th market: the higher its value, the more the transaction price must fluctuate in the opposite direction to flip its sign. For each configuration, we sample the absolute value of this statistics at the emergence of any triangular arbitrage opportunity and normalize its average by the initial center of mass $p_{\ell}(t_0)$, see Section \ref{appendix:calibration}, to make it comparable with the same quantity measured in other markets. For $\{+,+,+\}$ we find that  $\langle|\phi_{n,\ell}(t)|\rangle/p_{\ell}(t_0)$ is substantially higher for EUR/JPY than EUR/USD and USD/JPY, see Fig.~\ref{fig:trendstrengthhist}. As a result, predatory market orders are more likely to set the ground for transitions from $\{+,+,+\}$ to $\{-,+,+\}$ (35.8\%) or $\{+,-,+\}$ (33.6\%). Looking at these transitions on the opposite direction is even more compelling: $\{+,+,+\}$ is the most likely destination from both $\{-,+,+\}$ (37.5\%) and $\{+,-,+\}$ (36.4\%). This hints at the presence of a loop in which the ecology transits from $\{+,+,+\}$ to $\{-,+,+\}$ or $\{+,-,+\}$ and then moves back. Such dynamics find an explanation in the fact that the market that has recently flipped its state, causing a departure from $\{+,+,+\}$ towards $\{-,+,+\}$ or $\{+,-,+\}$, can be easily flipped back again before its \emph{resistance} to state changes $\phi_{n,\ell}(t)$ increases in absolute value. This happens when the arbitrager responds to a type 2 triangular arbitrage opportunity (i.e., $\mu^{II}(t) > 1$) when the ecology configuration is either $\{-,+,+\}$ or $\{+,-,+\}$. \\
The significant probabilities of returning to $\{+,+,+\}$ stem from the interplay of two elements. First, triangular arbitrage opportunities are more likely to be of type 2 than type 1 in both $\{-,+,+\}$ and $\{+,-,+\}$, see Fig.~\ref{fig:mubarplot}. Second, the markets with lowest \emph{resistance} to state changes $\langle|\phi_{n,\ell}(t)|\rangle/p_{\ell}(t_0)$ are EUR/USD for $\{-,+,+\}$ and USD/JPY  for $\{+,-,+\}$, see Fig.~\ref{fig:trendstrengthhist}, which are exactly the states that should be flipped to return to $\{+,+,+\}$.
The conditional transition probability matrix displayed in Table~\ref{tab:tranmatrix} reveals the presence of another configuration triplet (i.e., $\{-,-,-\}$, $\{-,+,-\}$ and $\{+,-,-\}$) exhibiting an analogous behavior while $\{-,-,+\}$ and $\{+,+,-\}$ are the only two configurations that are not part of any loop. Fig.~\ref{fig:clusteranalysis} shows this mechanism in action by displaying the sequence of ecology configurations during a segment of the model simulation. It is easy to observe how the system tends to move across configurations belonging to the same looping triplet for long, uninterrupted time windows. Ultimately, this peculiar mechanism increases, to different degrees, the appearance probabilities of configurations involved in these loops at the expenses of $\{-,-,+\}$ and $\{+,+,-\}$.
\\
To sum up, our model elucidates how
the interplay between different trading strategies entangles
the dynamics of different FX rates, leading to the characteristic shape of the cross-correlation 
functions observed in real trading data. The Arbitrager Model restricts
its focus to the interactions between two types of strategies, namely triangular arbitrage and
trend-following. Despite the simplicity of our framework, the interplay between these two strategies alone satisfactorily
reproduces the cross-correlation functions observed in real trading data. In particular,
trend-following strategies preserve the current combination of market states for some time
while reactions to triangular arbitrage opportunities \emph{influence} the behavior of trend-following
market makers by altering the price trend signals used in their dealing strategies. 
The interactions between these two strategies constantly push the system towards certain configurations
and away from others through multiple mechanisms. This can be easily seen in Fig.~\ref{fig:statusDuration} as two distinct statistics, the average expected lifetimes and the
appearance probability, put the eight configurations in the same order. For instance $\{+,+,+\}$ has the longer expected lifetime but also the highest appearance probability. This \emph{force} shapes the features of the statistical relationships between currency pairs. FX rates traded in markets that share the same state in configurations with higher (lower) appearance probabilities and longer (shorter) expected lifetimes are more likely to fluctuate in the same (opposite) direction. For instance, let us consider USD/JPY and EUR/JPY. These two markets have the same states in the four configurations with higher probabilities (i.e., $\{+,+,+\}$, $\{-,+,+\}$, $\{+,-,-\}$ and $\{-,-,-\}$) and opposite states in those with lower probabilities (i.e., $\{+,-,+\}$, $\{-,-,+\}$, $\{+,+,-\}$ and $\{-,+,-\}$). It follows that the probability of observing USD/JPY and EUR/JPY in the same state at a given point in time $t$ is $\approx$ 60\%, see Fig.~\ref{fig:Heterogenuity}. In these settings, the mid price dynamics of two FX rates become permanently entangled, leading to the cross-correlation functions displayed in Figs.~\ref{fig:TwoCorr}(b) and \ref{fig:Heterogenuity}.
\section{Discussion and outlook}\label{sec:final}
The purpose of this study was to obtain further insights on the microscopic origins of the widely documented cross-correlations among currencies. We took up this challenge by introducing a new ABM, the Arbitrager Model, in which market makers adopting trend-following strategies provide liquidity in three independent markets and interact with an arbitrager. In these settings, our model reproduced the characteristic shape of the cross-correlation function between between fluctuations of FX rate pairs under the assumption that triangular arbitrage is the only mechanism through which the different FX rates become synchronized. This suggests that triangular arbitrage plays a primary role in the entanglement of the dynamics of currency pairs in real FX markets. 
In addition, the model explains how the features of $\rho_{i,j}(\omega)$ emerges from the interplay between triangular arbitrage and trend-following strategies. In particular, triangular arbitrage \emph{influences} the trend-following behaviors of liquidity providers, driving the system towards certain combinations of price trend signs and away from others. This affects the probabilities of observing two FX rates drifting in the same or opposite direction, making one of the two scenarios more likely than the other. Ultimately, this entangles the dynamics of these prices, creating the significant cross-currency correlations that are reproduced in our model and observed in real trading data.
\\
The present study, finding a common ground between previous microscopic ABMs of the FX market and triangular arbitrage \cite{aiba2006microscopic, yamada2009solvable, kanazawa2018derivation}, sets a new benchmark for further investigations on the relationships between agent interactions and market interdependencies. In particular, it is the first ABM to provide a complete picture on the microscopic origins of cross-currency correlations. \\
The outcomes of this work open different research paths and raise new challenges that shall be considered in future studies. First, the Arbitrager Model could be further generalized by including a larger number of currencies, allowing traders to monitor different currency triangles. We assert that extending the number of available currencies could reveal new insights on i) statistical regularities related to the triangular arbitrage processes, such the distributions of $\mu^{I}(t)$ and $\mu^{II}(t)$, and ii) how the features of the cross-correlation function between two FX rates stem from a much more complex system in which the same FX rate is part of several triangles. Second, a potential extension of this model should consider the active presence of special agents operating in FX markets. For instance, simulating public interventions implemented by central banks could be a valuable exercise to understand how the large volumes moved by these entities affect the dynamics of the triangular arbitrage processes $\mu^{I}(t)$ and $\mu^{II}(t)$ and the local correlations (i.e., in the intervention time window) between currency pairs. 
Third, another interesting path leads to market design problems. In this study we have hypothesized a relationship between changes in the stabilization levels of the cross-correlation functions $\rho_{i,j}(\omega)$ and the different tick sizes adopted by EBS in the period covered by the employed dataset. Calling for further investigations, we believe that an extended version of the present model should examine how different tick sizes affect the correlations between FX rates.\\
We are confident that appropriate extensions and enhancements could turn the model into a valuable tool that could be used by exchanges, regulators and market designers. In particular, its simple settings would allow these entities to make predictions on how regulations or design changes could affect the relationships between FX rates and the properties (e.g., frequency, magnitude, duration, etc.) of triangular arbitrage opportunities in a given market. Furthermore, its applicability might attract the attention of other actors operating in the FX market, such as central banks. The ultimate objective of this work and its potential future extensions shall remain the provision of useful means to enhance the understanding of financial market dynamics, assisting the aforementioned entities in conceiving safer and more efficient trading environments. 
\section{Acknowledgments}
We thank EBS, NEX Group plc. for providing the data employed in this study. Alberto Ciacci acknowledges PhD studentships from
the Engineering and Physical Sciences Research Council through Grant No. EP/L015129/1. Alberto Ciacci and Takumi Sueshige thank Kiyoshi Kanazawa for fruitful discussions.
\section{Author Contributions}
\textbf{Conceptualization:} Hideki Takayasu, Misako Takayasu, Alberto Ciacci, Takumi Sueshige \\
\noindent
\textbf{Funding acquisition:} Alberto Ciacci, Takumi Sueshige, Misako Takayasu\\
\noindent
\textbf{Project supervision:} Hideki Takayasu, Misako Takayasu, Kim Christensen\\
\noindent
\textbf{Investigation:} Alberto Ciacci, Takumi Sueshige \\
\noindent
\textbf{Methodology:} Alberto Ciacci, Takumi Sueshige, Hideki Takayasu, Misako Takayasu \\
\noindent
\textbf{Software:} Takumi Sueshige, Alberto Ciacci \\
\noindent
\textbf{Visualization:} Takumi Sueshige \\
\noindent
\textbf{Validation:} Alberto Ciacci, Takumi Sueshige \\
\noindent
\textbf{Writing - original draft:} Alberto Ciacci, Takumi Sueshige \\
\noindent
\textbf{Writing - review and editing:} Alberto Ciacci, Takumi Sueshige, Hideki Takayasu, Misako Takayasu, Kim Christensen \\
\bibliography{paper}
\newpage
\appendix
\addcontentsline{toc}{section}{Appendices}
\setcounter{equation}{0}
\setcounter{figure}{0}
\setcounter{table}{0}
\setcounter{section}{0}
\renewcommand\theequation{S\arabic{equation}}
\renewcommand{\thetable}{S\arabic{table}}
\renewcommand\thefigure{S\arabic{figure}}   
\renewcommand\thesection{S\arabic{section}}
\section{Mid price patterns in the EBS market}\label{appendix:data1}
\begin{figure}[H]
  \begin{center}
    \includegraphics[width=13cm]{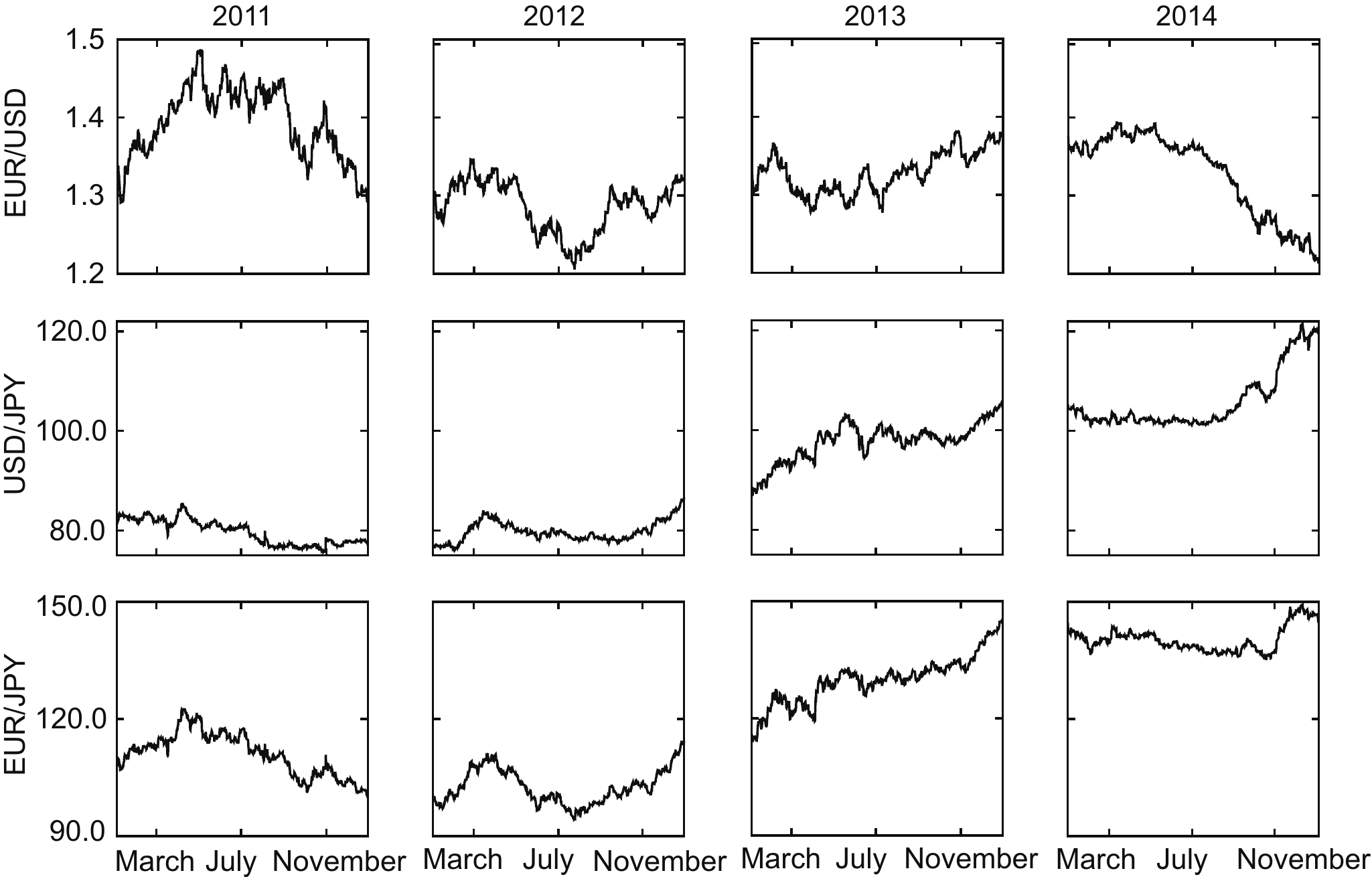}
    \caption{\textbf{Mid price patterns}. The panels show the dynamics of the mid price between January 1$^{\text{st}}$ 2011 and December 31$^{\text{st}}$ 2014 in the EUR/USD (top panels), USD/JPY (middle panels) and EUR/JPY (bottom panels) markets. Data is provided by EBS, see Section \ref{sec:data}.}
    \label{fig:PriceAll}
  \end{center}
\end{figure}
\section{Tick sizes adopted by EBS in the period 2011-2014}\label{appendix:data2}
\begin{table}[H]
 \centering
 \caption{\textbf{Tick sizes adopted in the EBS market.}}
 \begin{tabular}{|c|c|c|c|}
   \hline
   \shortstack{Initial Month} & \shortstack{USD/JPY} & \shortstack{EUR/USD} & \shortstack{EUR/JPY}\\
   \hline
   2011-03 & 0.01 & 0.0001 & 0.01\\
   2012-09 & 0.001 & 0.00001 & 0.001\\
   - & 0.005 & 0.00005 & 0.005\\
   \hline
 \end{tabular}%
  \caption*{
  \\
  EBS has changed the tick size twice in the period between January 1$^{\text{st}}$ 2011 and December 31$^{\text{st}}$ 2014. The table reports the initial month (yyyy-mm) of each implementation period (first column) and the corresponding tick size in the USD/JPY (second column), EUR/USD (third column) and EUR/JPY (fourth column) markets. See \cite{mahmoodzadeh2014tick} for further details.}
\label{tab:ticksizes}
\end{table}
\section{The Dealer Model (Yamada et al. 2009)}\label{appendix:yamadamodel}
The Dealer Model \cite{yamada2009solvable} introduces a simple market ecology in which $N$ agents interact in a single inter-dealer market where trading is organized in a LOB. For simplicity, the model assumes a continuous price grid, neglecting the role played by the tick size in real financial markets. Agents act as market makers by maintaining buy and sell limit orders through which they provide a bid and an ask quote to the market.
 \begin{figure}[H]
\centering
\includegraphics[scale = 0.15]{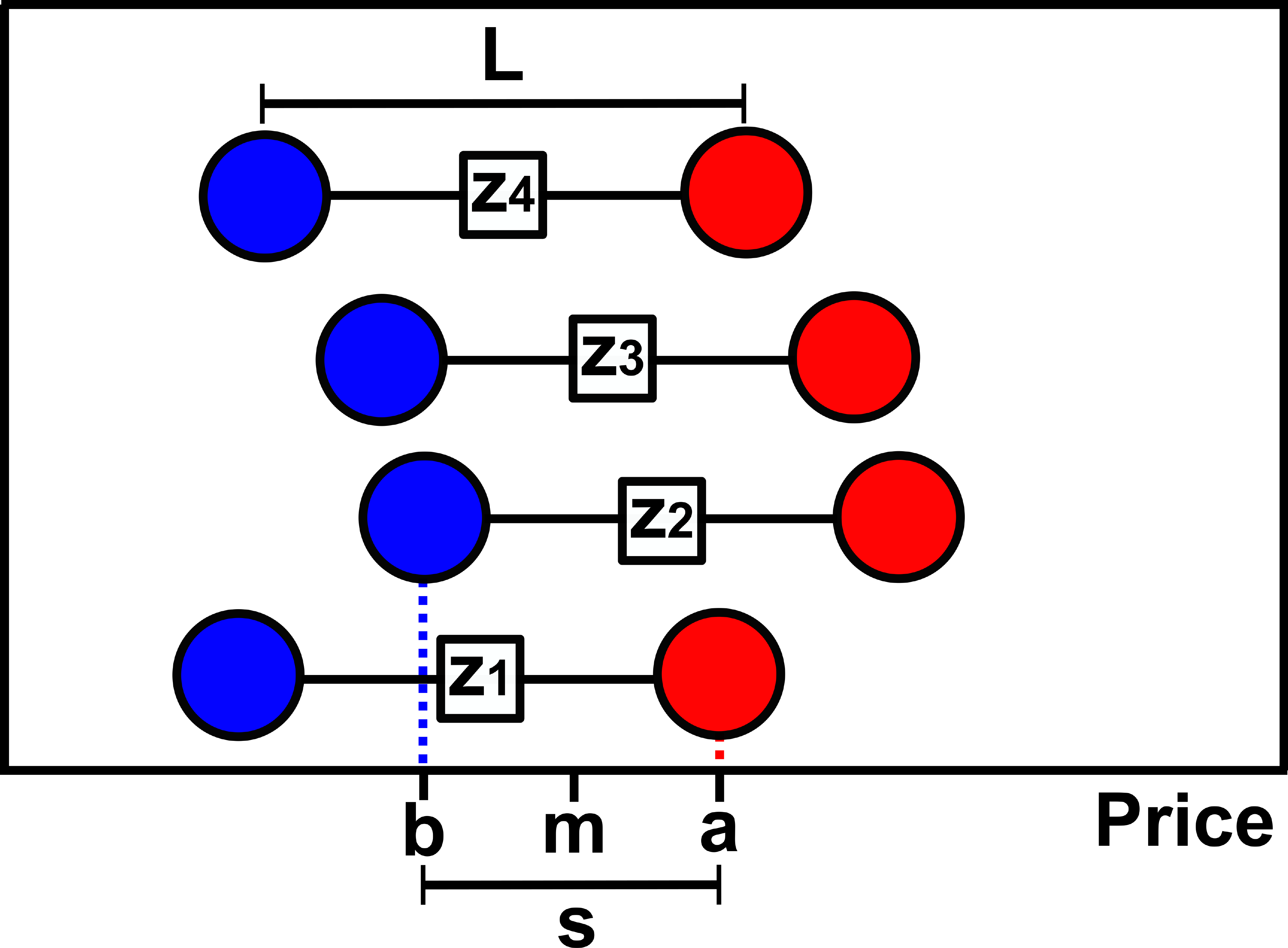}
\caption{\textbf{Dealer Model basics.} The market is participated by $N = 4$ market makers providing bid $b_{i}(t)$ (blue circles) and ask $a_{i}(t)$ (red circles) quotes, $i = 1,\dots,N$. The continuous price grid prevents limit orders to be queued at the same price level. Each limit order includes one unit of the traded currency. The market making spread $L = a_{i}(t) - b_{i}(t)$, $i = 1,\dots,N$ is the same for each agent and constant in time, that is, market makers only manage their dealing price 
$z_{i}(t) = (a_{i}(t) + b_{i}(t))/2$, $i = 1,\dots, N$ (white squares) to dynamically adjust their bid and ask quotes $b_{i}(t) = z_{i}(t) - L/2$ and $a_{i}(t) = z_{i}(t) + L/2$, $i = 1,\dots,N$. The vertical dashed lines mark the best bid $b(t)$ (blue) and ask $a(t)$ (red) quotes. The distance between the best quotes is the current spread $s(t) = a(t) - b(t)$. The current mid price is $m(t) = (a(t) + b(t))/2$.}
\label{yamadamodel}
\end{figure}
Transactions occur when the $i$-th market maker is willing to buy at a price that matches or exceeds the ask price of the $j$-th market maker (i.e., $b_{i} \geq a_{j}$). Trades are settled at the transaction price $p(g_t) = (a_{j}(t) + b_{i}(t))/2$, where $g_t$ is the number of transactions occurred in $[0,t[$. We stress that the mid-price $m(t)$ and the transaction price $p(g_t)$ are two different quantities. The former, being the mid point between the best quotes, is the \emph{center} of the LOB and can be tracked at any time step. The latter is sampled whenever two market makers engage in a trade.\\

\begin{figure}[H]
\centering
\includegraphics[width = 12.5 cm, height = 3 cm]{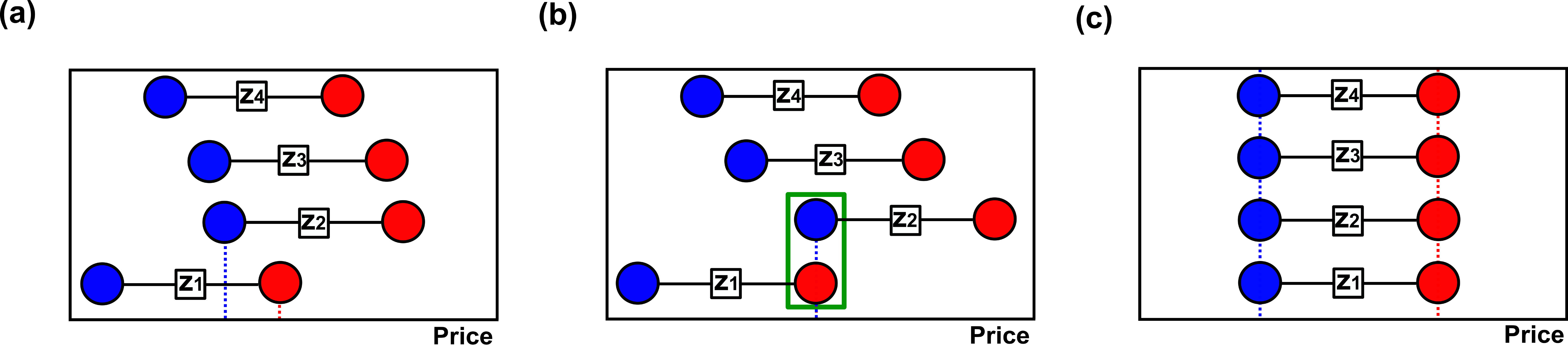}
\caption{\textbf{Interactions in the Dealer Model.} (a) Market makers are not engaging in transactions as the best bid price (blue dashed line) is smaller than the best ask price (red dashed line). (b) The best bid price matches the best ask price, prompting Market Makers \#1 and \#2 to exchange one unit of the traded FX rate (green box). The transaction price is the mid point between the two quotes $p = (a_1 + b_2)/2$. (c) This transaction prompts each market maker to update its dealing price to the latest transaction price (i.e., $z \rightarrow p$).}
\label{dealermodelinteractions}
\end{figure}

The Dealer Model assumes that a transaction prompts the entire market to immediately update their dealing prices $z_{i}(t + dt),\; i = 1,\dots,N$ to the latest transaction price $p(g_t)$, see Fig.~\ref{dealermodelinteractions}.
In the absence of interactions, market makers independently update their dealing prices by adopting a trend-based strategy
\begin{equation}\label{midpriceeq}
z_{i}(t) = 
 z_{i}(t - dt) + c\langle \Delta p\rangle_{n}dt + \sigma\sqrt{dt}\epsilon_{i}(t) \;\;\; i = 1,\dots, N
\end{equation}
where $\sigma > 0$ and $\epsilon_{i}(t) \sim \mathcal{N}(0, 1)$. The term 
\begin{equation}\label{trend}
\langle \Delta p\rangle_{n} = \frac{2}{n(n+1)}\sum\limits_{k = 0}^{n-1}(n - k)(p(g_t - k) - p(g_t - k - 1))
\end{equation}
is a weighted average of the last $n < g_t$ changes in the transaction price $p$.
\\
The real-valued parameter $c$ controls how the current price trend $\langle\Delta p\rangle_{n}$ influences market makers' strategies. For instance, $c > 0$ represents a market maker that tends to adjust its dealing price $z(t)$ in the direction of the price trend (i.e., trend-following). Conversely, $c < 0$ characterizes a market maker that tends to adjust its dealing price in the opposite direction of the price trend (i.e., contrarian).
\section{Agent interactions in the Arbitrager Model}\label{appendix:model1}
\begin{figure}[H]
\centering
\includegraphics[width = 12.5 cm, height = 3.0 cm]{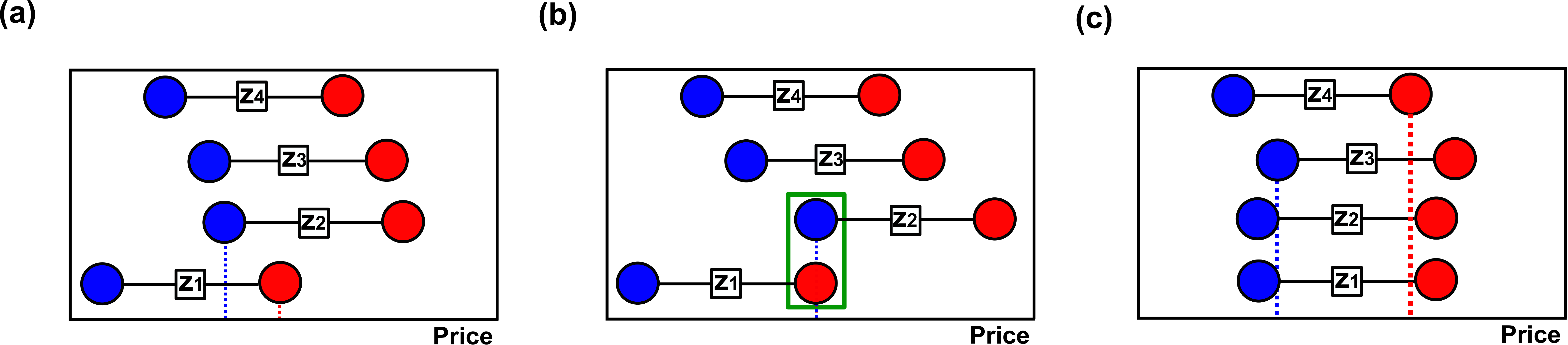}
\caption{\textbf{Interactions in the Arbitrager Model.} (a) Market makers are not engaging in transactions as the best bid price (blue dashed line) is smaller than the best ask price (red dashed line). (b) The best bid price matches the best ask price, prompting Market Makers \#1 and \#2 to exchange one unit of the traded FX rate (green box). The transaction price is the mid point between the two quotes $p = (a_1 + b_2)/2$. (c) This transaction prompts the two transacting market makers to re-adjust their dealing prices $z$ to the latest transaction price $p$.}
\label{arbitragermodelinteractions}
\end{figure}

\begin{figure}[H]
\centering
\includegraphics[width = 13cm, height = 9.5cm]{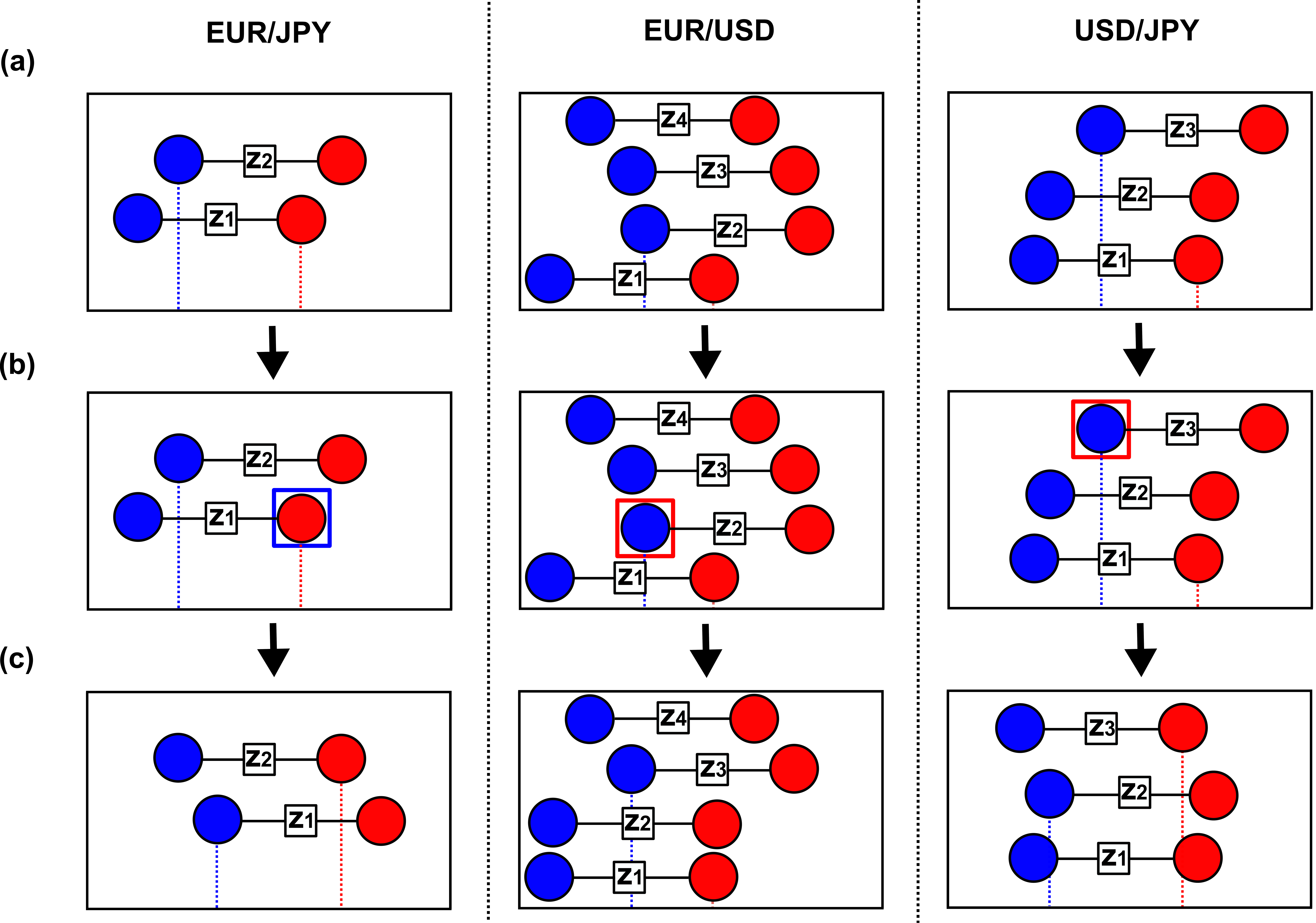}
\caption{\textbf{Exploiting a triangular arbitrage opportunity in the Arbitrager Model.} (a) The states of the three markets before the emergence of an exploitable triangular arbitrage opportunity. (b) When $\mu^{I}(t) \geq  1 $, the arbitrager submits a buy market order (blue square) in the EUR/JPY market and sell market orders (red squares) in the EUR/USD and USD/JPY markets, matching the encapsulated limit orders (i.e., Market Maker \# 1 in EUR/JPY, Market Maker \# 2 in EUR/USD and Market Maker \# 3 in USD/JPY). (c) The transacting market makers re-adjust their dealing prices to the quote matched by the arbitrager's market order (e.g., $z_{1,\text{EUR/JPY}}(t + dt)\rightarrow a_{1,\text{EUR/JPY}}(t)$), causing a mid price change in each market.}
\label{fig:arbitragermodelattack}
\end{figure}

In this section we provide further details on the mechanisms ruling agents interactions in the Arbitrager Model, see Figs.~\ref{arbitragermodelinteractions} and \ref{fig:arbitragermodelattack}.
\section{Initialization and dynamic control of the Arbitrager Model}\label{appendix:calibration}
\subsection{Introduction}
Kanazawa et al. \cite{kanazawa2018derivation} have recently introduced a microscopic model of the interactions between high frequency traders (HFTs) and investigated its theoretical aspects by adapting Boltzmann and Langevin equations to this specific context. The results of this work have been further formalized in a parallel study from the same authors \cite{kanazawa2018kinetic}.  The dealing price updates in the HFT model are driven by the following dynamics
\begin{equation}\label{midpriceeq3}
z_{i,\ell}(t) = z_{i,\ell}(t - dt) + c_{\ell}^{*}\text{tanh}\left(\frac{p_{\ell}(g_{t,\ell}) - p_{\ell}(g_{t,\ell} - 1) }{\Delta p^{*}_{\ell}}\right)dt + \sigma_{\ell}\sqrt{dt}\epsilon_{i,\ell}(t) \;\;\; i = 1,\dots, N_{\ell}
\end{equation}
where $\Delta p^{*}_{\ell}$, $c_{\ell}^{*}$ are constants while the other variables and constants have the same meaning as in Eq.~\eqref{midpriceeq2}. It can be shown that setting $\Delta p^{*}_{\ell} \gg p_{\ell}(g_{t\ell}) - p_{\ell}(g_{t\ell} - 1)$ allows for a linear approximation of
Eq.~\eqref{midpriceeq3} that resembles the dynamics of the dealing price updates in the Arbitrager Model, see Eq.~\eqref{midpriceeq2}. This correspondence allows us to exploit the theoretical results of \cite{kanazawa2018derivation, kanazawa2018kinetic} to achieve a satisfactory control of the dynamics of the Arbitrager Model. For instance, Fig.~\ref{fig:TimeDifferencesModel} shows that the average time between consecutive transactions in simulations of the Arbitrager Model is in strong agreement with its theoretical value estimated in the framework of Kanazawa et al. \cite{kanazawa2018derivation,kanazawa2018kinetic}. In the following sections we provide details on how the parameters governing the evolution of our model have been set in the simulations discussed in this study.
\subsection{Overview of the model parameters}
\begin{table}[H]
 \centering
 \caption{\textbf{Parameters governing the dynamics of the Arbitrager Model}}
 \begin{tabular}{|c|c|c|c|}
   \hline
   \textbf{Name} & \textbf{Symbol} & \textbf{Dimension} & \textbf{Section}\\
   \hline
   Initial center of mass & $p_{\ell}(t_0)$ & price & \ref{appendix:initiallob} \\
   Market making spread &$L_{\ell}$ & price & \ref{appendix:initiallob}\\
   Number of market participants &$N_{\ell}$ & dimensionless & \ref{appendix:Nsigmagamma}\\
   Volatility of dealing price updates &$\sigma_{\ell}$ & price/$\sqrt{\text{time}}$ & \ref{appendix:Nsigmagamma}\\
   Average time between transactions &$\Gamma$ & time (sec) & \ref{appendix:Nsigmagamma}\\
   Discretized time step &$\Delta t$ & time (sec) & \ref{appendix:Nsigmagamma}\\
   Price changes accounted in $\phi_{n,\ell}(t)$ &$n$ & dimensionless & \ref{appendix:xiandn}\\
   Scaling of the weight function in $\phi_{n,\ell}(t)$ &$\xi$ & dimensionless & \ref{appendix:xiandn}\\
   Trend-following strength &$c_{\ell}$ & price/time & \ref{appendix:c}\\
   \hline
 \end{tabular}%
  \caption*{
  \\
  The evolution of the Arbitrager Model ecology is controlled by 9 parameters. For each parameter, we report its nomenclature, symbol, dimension and the section in which we explain how its value is set in the simulations presented in Fig.~\ref{fig:TwoCorr}.}
\label{tab:modelpar}
\end{table}
\subsection{Initial state of the LOB}\label{appendix:initiallob}
To initialize the $\ell$-th LOB, we first fix its initial center of mass $p_{\ell}(t_0)$ and the constant market making spread $L_{\ell}$. The former is set arbitrarily to a value with the same magnitude of the mid-price patterns observed in real trading data, see Fig.~\ref{fig:PriceAll}. Following the analysis of \cite{kanazawa2018derivation}, we fix the market making spread in the USD/JPY market to $L_{\text{USD/JPY}} = 0.05$. For simplicity, the market making spread in other markets is set such that it becomes proportional to the size of $p_{\ell}(t_0)$, that is $L_{\ell} = L_{\text{USD/JPY}}\times(p_{\ell}(t_0)/p_{\text{USD/JPY}}(t_0))$.
\begin{table}[H]
	\centering
	\caption{\textbf{Initial center of mass and market making spread in each market}}
	\begin{tabular}{|c|c|c|}
		\hline
		\textbf{Exchange Rate} & $\boldsymbol{p(t_0)}$ & $\boldsymbol{L}$ \\ \hline
		EUR/USD              & 1.25            & 0.05
		   \\ \hline
		USD/JPY               & 110            & 0.05$\times$(110/1.25)     \\ \hline
		EUR/JPY               & 137.5            & 0.05$\times$(137.5/1.25)           \\ \hline
	\end{tabular}
	\caption*{Values of $p_{\ell}(t_0)$ and $L_{\ell}$ for EUR/USD, USD/JPY and EUR/JPY.}
	\label{tab:initialpar}
\end{table}
At this point we use the values in Table~\ref{tab:initialpar} to obtain the initial dealing prices for each trader and market, thus revealing the initial profile of the LOBs
\begin{equation}\label{lambda}
z_{i,\ell}(t_0) = \begin{cases}
\frac{L_{\ell}}{2}\left(\sqrt{2u_{i,\ell}}-1\right)+p_{\ell}(t_0), & \text{if } 0 < u_{i,\ell} \le 0.5\\
\frac{L_{\ell}}{2}\left(1-\sqrt{2(1-u_{i,\ell})}\right)+p_{\ell}(t_0)  & \text{otherwise}
\end{cases}
\end{equation}
where $u_{i,\ell} \sim U(0,1)$ is an uniformly distributed random variable. The expression in Eq.~\eqref{lambda} is derived from the inverse function sampling procedure. Let $r \equiv (z(t_0) - p(t_0))$ be the relative distance between an initial dealing price $z(t_0)$ and the initial center of mass price $p(t_0)$. The LOB profile is stable when the probability density function (PDF) of $r$ is
\begin{equation}
\psi_{\ell}(r) = \begin{cases}
\frac{2}{L_{\ell}}\left(1 - |\frac{2r}{L_{\ell}}|\right) & \text{if } |r| \leq \frac{L_{\ell}}{2} \\
0 & \text{otherwise}
\end{cases}
\end{equation}
it follows that the cumulative density function (CDF) of $r$ is 
\begin{equation}\label{CDF}
\Psi_{\ell}(r) = \begin{cases}
\frac{1}{2L^{2}_{\ell}}\left(L_{\ell}+2r\right)^2, & \text{if } -\frac{L_{\ell}}{2} < r \le 0\\
-\frac{1}{2L^{2}_{\ell}}\left(L_{\ell}-2r\right)^2+1  & \text{otherwise}
\end{cases}
\end{equation}
then, we compute the inverse function of Eq.~\eqref{CDF}
\begin{equation}\label{ICDF}
\Psi_{\ell}^{-1}(y)= \begin{cases}
\frac{L_{\ell}}{2}\left(\sqrt{2y}-1\right), & \text{if } 0 < y \le 0.5\\
\frac{L_{\ell}}{2}\left(1-\sqrt{2(1-y)}\right)  & \text{if } 0.5 < y \le 1
\end{cases}
\end{equation}
Finally, assuming that $y = u \sim U(0,1)$, we use Eq.~\eqref{ICDF} to obtain the value of the initial dealing price $z_{i,\ell}(t_0)$, see Eq.~\eqref{lambda}.
\begin{figure}[H]
	\begin{center}
		\includegraphics[width=13cm]{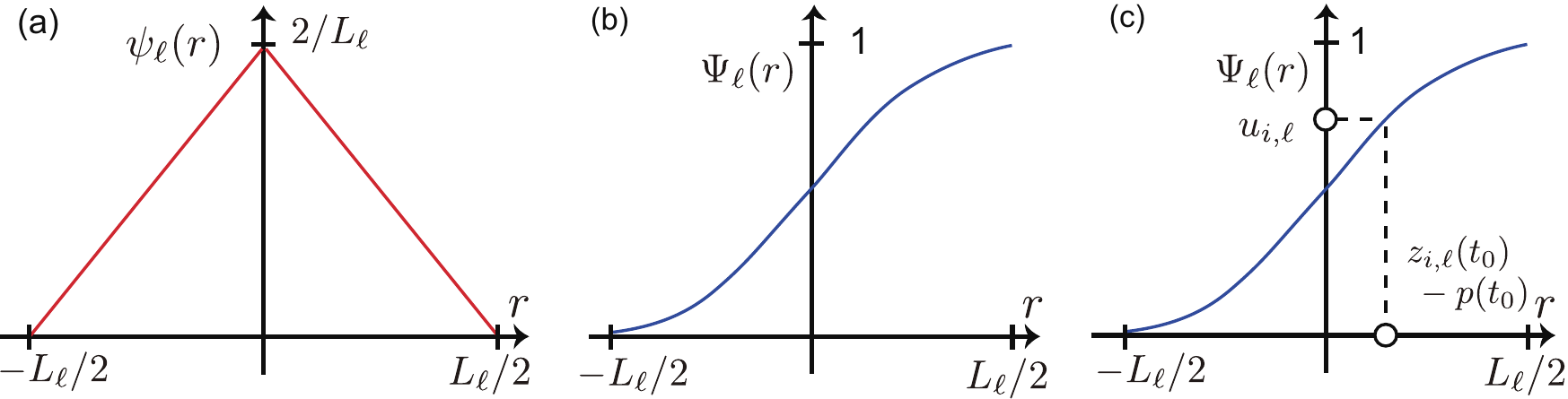}
		\caption{\textbf{Inverse function sampling in the context of the Arbitrager Model.} (a) The LOB profile is stable if the PDF of $r$ corresponds to the triangular function $\psi_\ell(r)$ \cite{kanazawa2018kinetic}. (b) The CDF $\Psi_\ell(r)$. (c) Schematic of the inverse function sampling applied to $\Psi_{\ell}(r)$.		}
		\label{fig:InitLimPrice}
	\end{center}
\end{figure}
\subsection{Relationships between simulation time and real time}\label{appendix:Nsigmagamma}
\begin{figure}[H]
	\begin{center}
		\includegraphics[width=9cm]{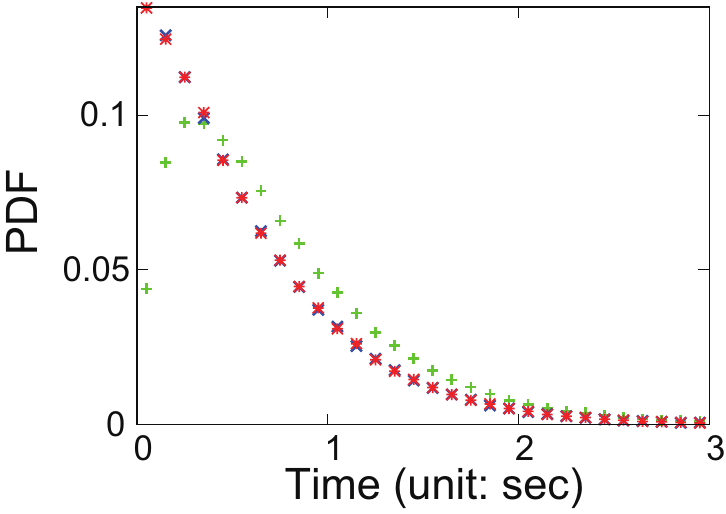}
		\caption{\textbf{Waiting times statistics in the Arbitrager Model.} Probability density functions (PDFs) of the waiting times between consecutive transactions for EUR/USD (red), USD/JPY (blue) and EUR/JPY (green) in the Arbitrager Model. Simulations are performed under the same settings of the experiment presented in Fig.~\ref{fig:TwoCorr}(b), bottom panel. For an adequate comparison against the theoretical predictions of the average time between consecutive transactions \cite{kanazawa2018kinetic}, the PDFs do not account for transactions triggered by the arbitrager.
		}
		\label{fig:TimeDifferencesModel}
	\end{center}
\end{figure}

Kanazawa et al. \cite{kanazawa2018kinetic} found that the average time between two consecutive transactions is
        \begin{equation}\label{theoreticalgamma}
        \Gamma = \frac{L^{2}_{\ell}}{2N_{\ell}\sigma^{2}_{\ell}},
        \end{equation}
In Section \ref{appendix:initiallob} we have fixed $L_{\ell}$. For the sake of simplicity, we assume that the three markets in the Arbitrager Model \emph{moves at the same pace}, on average. This implies that $\Gamma$ is the same in each market and constant in time. Informed by real trading data, we fix $\Gamma$ as follows. We consider each market separately and calculate the average waiting times between consecutive transactions in each trading year. This leads to 12 averages (i.e., 4 years $\times$ 3 FX rates). Finally, we compute the median of these averages to obtain a common value for $\Gamma \approx 0.7$ sec.
\begin{figure}[H]
	\begin{center}
		\includegraphics[width=13cm]{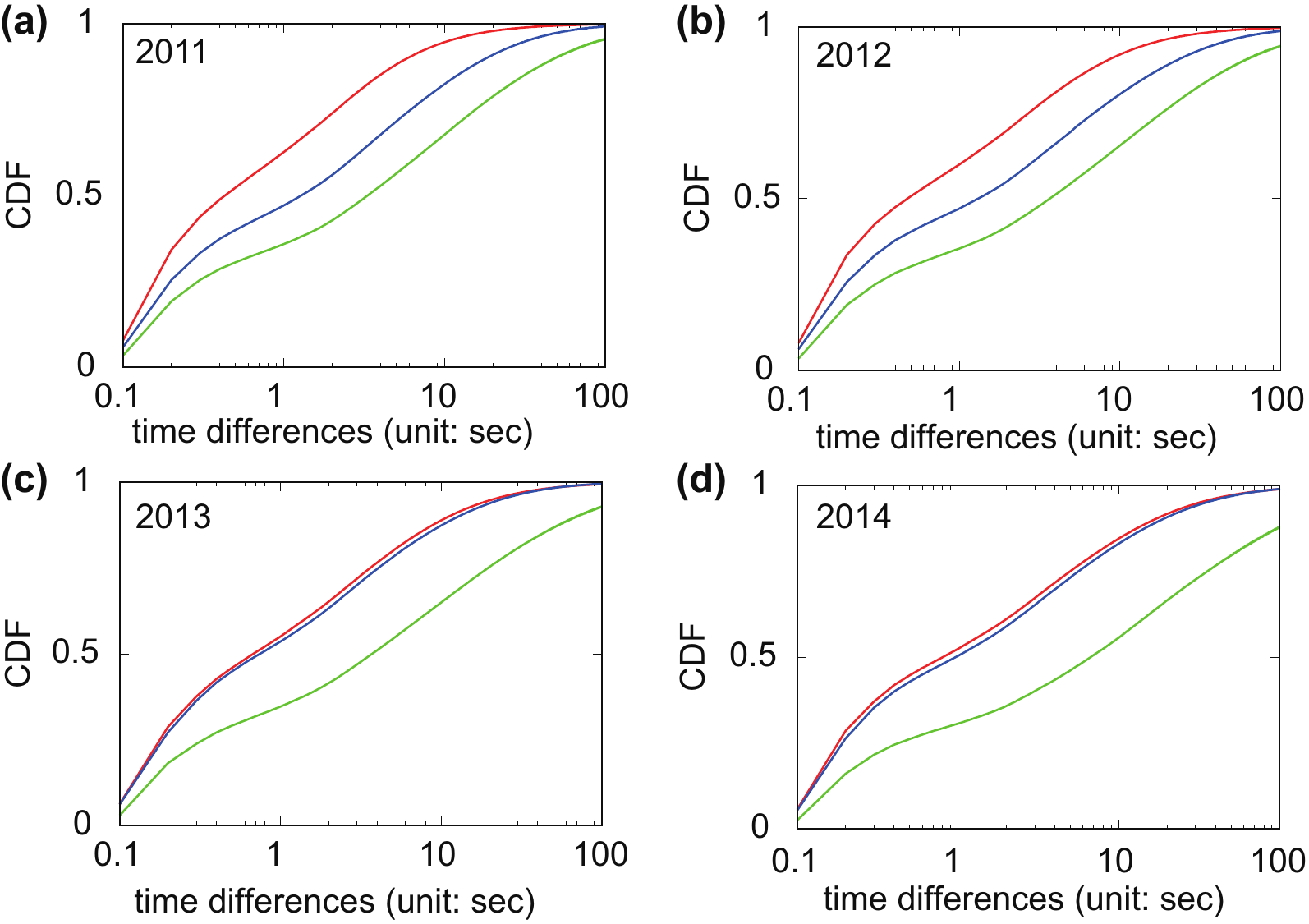}
		\caption{\textbf{Waiting times statistics in real trading data.} Cumulative density functions (CDFs) of the waiting times between consecutive transactions for EUR/USD (red), USD/JPY (blue) and EUR/JPY (green) in (a) 2011, (b) 2012, (c) 2013 and (d) 2014. Data is provided by EBS, see Section \ref{sec:data}.
		}
		\label{fig:TimeDifferencesReal}
	\end{center}
\end{figure}
To ensure that simulations of our model maintain $\Gamma \approx 0.7$ sec, we must fix $N_{\ell}$ and $\sigma_{\ell}$ such that Eq.~\eqref{theoreticalgamma} is satisfied. The number of market makers participating each market is set heuristically by considering several combinations ($N_{\text{EUR/USD}}$,$N_{\text{USD/JPY}}$,$N_{\text{EUR/JPY}}$) and examining how well the model-based cross-correlation function $\rho_{i,j}(\omega)$ replicates the same function built on real trading data.
Having fixed $N_{\ell}$, the volatility of the dealing price updates is found by rearranging Eq.~\eqref{theoreticalgamma}
\begin{equation}
    \sigma_{\ell} = \frac{L_{\ell}}{\sqrt{2N_{\ell}\Gamma}}.
\end{equation}
Finally, the amplitude of a discretized time step in the model simulation $\Delta t$ should be set such that $\Delta t \ll \Gamma$. We fix $\Delta t = 0.01$ sec and use this constant in the discrete approximation of Eq.~\eqref{midpriceeq2}. 
\begin{table}[H]
	\centering
	\caption{\textbf{Approximate equivalences between real and model time}}
	\begin{tabular}{|c|c|}
		\hline
		\textbf{sec} &  \textbf{time steps} \\ \hline
        1 & 100\\ \hline
        10 & 1000\\ \hline
        60 & 6000\\ \hline
	\end{tabular}
	\caption*{Using the results of Kanazawa et al. \cite{kanazawa2018derivation,kanazawa2018kinetic}, we establish an approximate equivalence between real and model time. Assuming $\Delta t = 0.01$ sec, we show how many time steps roughly equate to 1 sec, 10 sec and 1 min.}
	\label{tab:timeconversion}
\end{table}
Fig.~\ref{fig:TimeDifferencesModel} shows the distributions of the time between consecutive transactions in simulations of the Arbitrager Model. We find that the average waiting time in our simulations is $\tilde{\Gamma} \approx 0.65$ sec (65 time steps), which is very close to the theoretical value $\Gamma \approx 0.7$ sec. Acknowledging the simplifications that characterize our model (e.g., $\Gamma$ is the same in each market), we use the approximate equivalence $\Delta t = 0.01$ sec to convert simulation time steps in real time and compare the stabilization of the data-based and model-based cross-correlation functions $\rho_{i,j}(\omega)$, see Figs.~\ref{fig:TwoCorr} and \ref{fig:TwoCorr2}
\subsection{Parameters involved in the calculation of the current price trend}\label{appendix:xiandn}
The calculation of the price trend process $\phi_{n,\ell}(t)$, see Eq.~\eqref{trend2}, requires us to set the number of accounted transaction price changes $n$ and the scaling constant of the exponential weighting function $\xi$. In the simulations presented in Figs.~\ref{fig:TwoCorr} and \ref{fig:TwoCorr2} we arbitrarily set $n = 15$ observations and $\xi = 5$. These choices allow us to model a scenario in which trend-following market makers do not exclusively rely on the latest change in the transaction price to determine the current \emph{direction} of the market. Instead, they compute a weighted average of the most recent price changes where weights are calculated according to an exponential function.
\subsection{Trend-following strength parameter}\label{appendix:c}
The trend-following strength parameter $c$ determines how the sign and value of the current price trend $\phi_{n,\ell}(t)$ affect the strategic decisions of the participating market makers. When $c > 0$, market makers are likely to update their dealing prices $z(t)$ upward when the price trend is positive and downward when the price trend is negative. Conversely, $c < 0$ indicates that market makers are more likely to update their dealing prices in the opposite direction of the price trend sign.\\
Recently, Sueshige \textit{et al.} \cite{sueshige2018ecology} have classified the strategic behavior of FX traders by examining EBS data covering the trading activity in the USD/JPY market during the week starting from June $5^{\text{th}}$ 2016. They found that a significant fraction of traders adopt trend-following strategies (i.e., $c > 0$). This observation is consistent with the model of Yura \textit{et al.} \cite{yura2014financial}. \\
Relying on these studies, we enforce the assumption that market makers populating the Arbitrager Model ecology adopt trend-following strategies (i.e., $c > 0$). For simplicity, we assume that $c$ is the same for every market maker and across markets. To fix $c$, we use Eq. (91) in \cite{kanazawa2018kinetic}. The nondimensional parameter 
\begin{equation}
\Delta \tilde{p}^* = 1/(c\Gamma)
\end{equation}
shall take values that are not far from 2 for the model to produce the marginal trend-following behavior, which successfully replicated various statistical properties of real trading data in \cite{kanazawa2018kinetic}. This motivates us to set $c = 0.8$, thus obtaining $\Delta \tilde{p}^* \approx 1.79$.
\section{An extended version of the Arbitrager Model}\label{appendix:extended}
\subsection{Motivations}
The Arbitrager Model qualitatively replicates the shape of the cross-correlation functions $\rho_{i,j}(\omega)$ and provides important insights on how the microscopic interactions between market makers and arbitragers entangles the dynamics of different FX rates. However, the cross-correlation functions $\rho_{i,j}(\omega)$ reproduced by this extremely simple model present two features that are not found in real trading data. First, on extremely short time-scales (i.e., $\omega \rightarrow 0$ sec) the model-based $\rho_{i,j}(\omega)$ does not approach zero as the same function built on real trading data. Second, the model-based $\rho_{i,j}(\omega)$ flattens when $\omega \gtrapprox 30$ sec while the data-based $\rho_{i,j}(\omega)$ flattens when $\omega \gtrapprox 10$ sec.
\begin{figure}[H]
	\begin{center}
		\includegraphics[width=13cm]{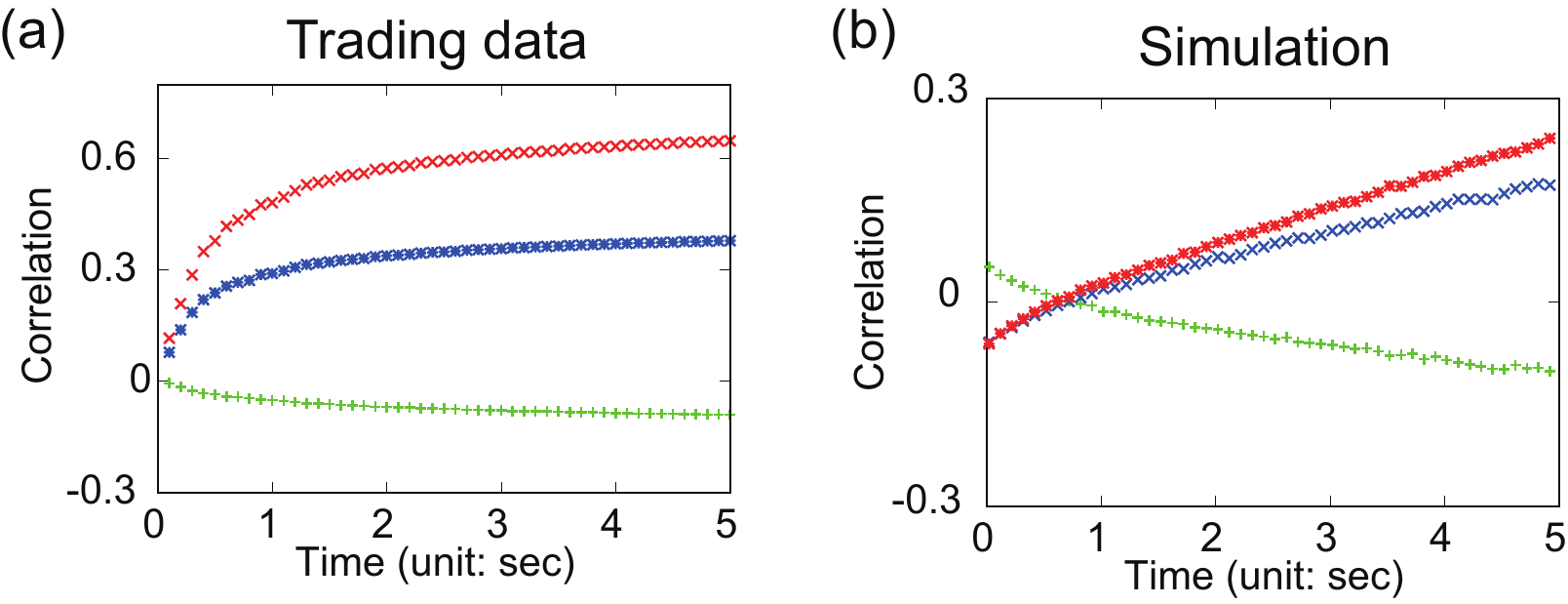}
		\caption{\textbf{Trading data vs. model based cross-correlations functions.} Enlarged visualization of the cross-correlation functions $\rho_{i,j}(\omega)$ presented in Fig.~\ref{fig:TwoCorr}. (a) Real market data (EBS) in 2013. (b) Arbitrager Model simulations. Contrarily to the cross-correlation functions displayed in (a), the model-based $\rho_{i,j}(\omega)$ takes non-zero values when $\omega \rightarrow 0$ sec and stabilizes on longer time-scales. Simulations are performed under the same settings of the experiment presented in Fig.~\ref{fig:TwoCorr}(b), bottom panel.
		Details on the initialization of the model and the conversion between simulation time (i.e., time steps) and real time (i.e., sec) are provided in Section \ref{appendix:calibration}.}
		\label{fig:EnlargedrossCorr}
	\end{center}
\end{figure}
We assert that the differences analyzed in Fig.~\ref{fig:EnlargedrossCorr} stem from the extreme simplicity of the Arbitrager Model. To verify this assertion, we introduce and examine the behavior of a modified version of the model which mimics more features of real FX markets. This extended, more realistic framework retains the same fundamental rules of the Arbitrager Model, that is, i) market makers continuously provide liquidity in a single market and ii) the arbitrager is the only agent allowed to operate across markets through the submission of predatory market orders. However, it also adds three distinct features inspired by real markets practices. First, agents' responses to triangular arbitrage opportunities emerge from a more rational decision making process in which they take into the account the risks associated to the implementation of this strategy. Second, market makers foresee predatory market orders and re-adjust their quotes in advance, reducing the likelihood of being matched by arbitragers' orders. Third, market makers operating in the EUR/JPY market peg their quotes to the implied best bid and ask prices with probability $\gamma$. This introduces an additional toy (i.e., unrealistic) mechanism through which the dynamics of different FX rates become entangled.
\subsection{A more realistic decision making process}
\subsubsection{The arbitrager}
In the original model, see Section \ref{sec:Model}, the arbitrager automatically submits predatory market orders as soon as Eqs.~\eqref{dmarb1} or \eqref{dmarb2} exceeds the unit. In real FX markets this decision is far less trivial as these orders might not be executed at the prices used in the calculation of Eqs.~\eqref{dmarb1} and \eqref{dmarb2}. For instance, faster traders could have already exploited the existing opportunity, pushing back Eqs.~\eqref{dmarb1} or \eqref{dmarb2} below the unit. As a result, the profitable misprice evaporates, exposing slower arbitragers to the risk of generating losses. We introduce a more realistic decision making process in which the arbitrager takes into the account the risks associated with this trading strategy. In particular, the arbitrager submits market orders if one of the following conditions is satisfied
\begin{subequations}
\begin{equation}\label{muthreshold1}
	\mu^{I}(t) \geq   1 + \zeta_{\text{A}}(t)
\end{equation}
\begin{equation}\label{muthreshold2}
	\mu^{II}(t) \geq   1 + \zeta_{\text{A}}(t)
\end{equation}
\end{subequations}
where $\zeta_{\text{A}}(t) \sim \text{exp}(\lambda_{\text{A}})$. The parameter $\lambda_{\text{A}}$ represents the risk profile of the arbitrager. The higher the value of $\lambda_{\text{A}}$, the more profitable the gap between real and implied prices must be to \emph{convince} the arbitrager to exploit the current opportunity.
\subsubsection{Market makers}
The submission of predatory market orders ensures immediate execution, forcing the matched market makers to either sell \emph{too low} or buy \emph{too high}. In the original Arbitrager Model, see Section \ref{sec:Model}, market makers remain indifferent to triangular arbitrage opportunities, that is, they do not attempt to anticipate the arbitrager to avoid predatory market orders. However, it is plausible that such a simplifying assumption does not adequately describe the behavior of liquidity providers acting in real FX markets. In this extension of the Arbitrager Model we enhance the strategic behaviors of market makers by allowing them to foresee the arbitrager's moves and re-adjust their quotes accordingly. Their dealing price updates are driven by Eq.~\eqref{midpriceeq2}, however, they also track the likelihood of engaging in an unfavourable transaction with the arbitrager. For instance, the $i$-th market maker operating in the EUR/JPY market monitors its exposure to predatory market orders by calculating the following ratios 
\begin{subequations}
	\begin{equation}\label{dmchi1}
	\chi_{i,\text{EUR/JPY}}^{I}(t) = \frac{b_{\text{USD/JPY}}(t)\times b_{\text{EUR/USD}}(t)}{a_{i,\text{EUR/JPY}}(t)} 
	\end{equation}
	\begin{equation}\label{dmchi2}
	\chi_{i,\text{EUR/JPY}}^{II}(t) = \frac{b_{i,\text{EUR/JPY}}(t)}{a_{\text{USD/JPY}}(t)\times a_{\text{EUR/USD}}(t)}
	\end{equation}
\end{subequations}
where $b_{i,\ell}(t)$ and $a_{i,\ell}(t)$ are the current bid and ask limit prices of the $i$-th market maker and $b_{\ell}(t)$ and $a_{\ell}(t)$ are the current best quotes in the $\ell$-th market. Clearly, Eqs.~\eqref{dmchi1} and \eqref{dmchi2} can be straightforwardly rewritten for market makers operating in the USD/JPY or EUR/USD markets.\\
The more Eqs.~\eqref{dmchi1} or \eqref{dmchi2} exceeds the unit, the larger the discrepancy between the current quote of the $i$-th market maker and the implied best cross FX rate. In the former case, the $i$-th market maker is underpricing EUR/JPY, facing the risk of selling \emph{too low}. In the latter case, the $i$-th market maker is overpricing EUR/JPY, facing the risk of buying \emph{too high}. As the implied cross FX rate is the same for every agent, the market maker with the highest value of $\chi$ is always the one who is offering the best quote, hence the first to be matched by predatory market orders.\\
In the same spirit of Eqs.~\eqref{muthreshold1} and \eqref{muthreshold2}, we assume that the $i$-th market maker, perceiving a high risk of interacting with the arbitrager, deletes and re-adjusts its current quotes if one of the following conditions is satisfied
\begin{subequations}
\begin{equation}\label{chithreshold1}
	\chi_{i,\text{EUR/JPY}}^{I}(t) \geq   1 + \zeta_{\text{MM, EUR/JPY}}(t)
\end{equation}
\begin{equation}\label{chithreshold2}
	\chi_{i,\text{EUR/JPY}}^{II}(t) \geq   1 + \zeta_{\text{MM, EUR/JPY}}(t)
\end{equation}
\end{subequations}
where $\zeta_{\text{MM, EUR/JPY}}(t) \sim \text{exp}(\lambda_{\text{MM, EUR/JPY}})$. The parameter $\lambda_{\text{MM},\text{EUR/JPY}}$ represents the average risk profile (i.e., is the same for every market maker) in the EUR/JPY market. The lower the value of $\lambda_{\text{MM},\text{EUR/JPY}}$, the less market makers tolerate their exposure to predatory market orders. 
\\
When Eqs.~\eqref{chithreshold1} or \eqref{chithreshold2} is satisfied, the $i$-th market maker sets its dealing price to the current mid price $m_{\text{EUR/JPY}}(t) = (a_{\text{EUR/JPY}}(t) + b_{\text{EUR/JPY}}(t))/2$, rejecting the update imposed by Eq.~\eqref{midpriceeq2} to reduce the risk of engaging in a transaction with the arbitrager. This mimics real traders deleting their limit orders queued in the very first levels of the LOB to replace them with new orders lying farther away from the current best quotes. 
\begin{figure}[H]
\centering
\includegraphics[scale=0.115]{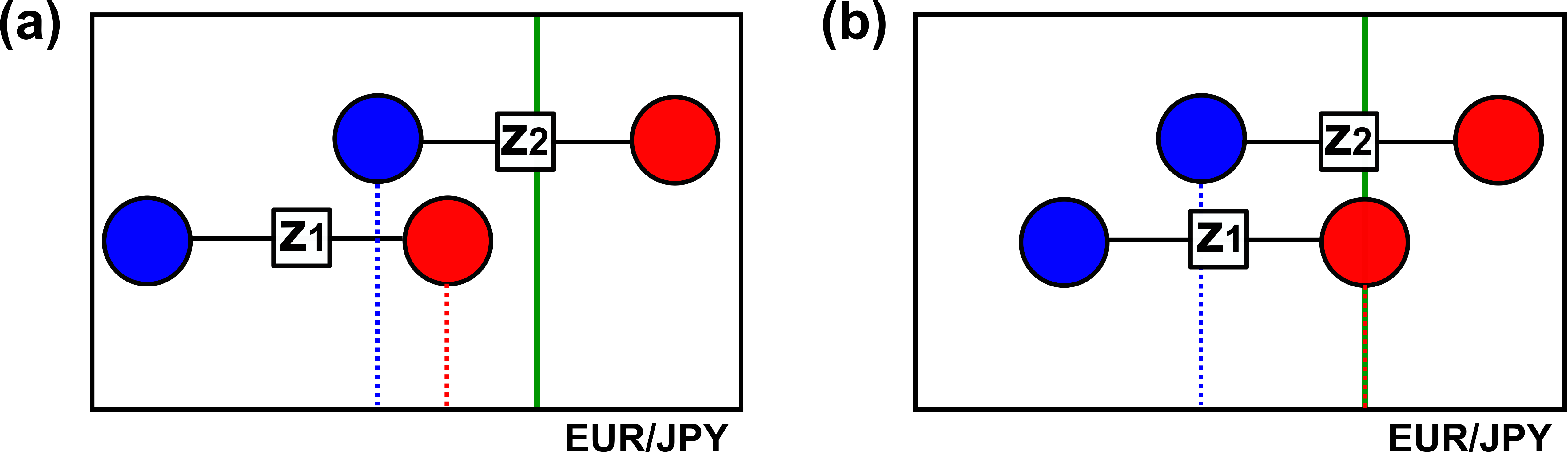}
    \caption{\textbf{Foreseeing a triangular arbitrage opportunity in the Arbitrager Model.} The plot considers the EUR/JPY market. The best bid and ask quotes are marked by the blue and red dashed lines, respectively. The implied best bid price of EUR/JPY (i.e., $b_{\text{USD/JPY}}\times b_{\text{EUR/USD}}$) is denoted by the green solid line. (a) The current best ask quote $a_{\text{EUR/JPY}}$ (red dashed line) is smaller than the implied best bid quote $b_{\text{USD/JPY}}\times b_{\text{EUR/USD}}$ (green solid line). This misprice exposes Market Maker \#1 to the risk of transacting with the arbitrage who wants to \emph{buy low} $a_{\text{EUR/JPY}}$ and \emph{sell high} $b_{\text{USD/JPY}}\times b_{\text{EUR/USD}}$. (b) When $\chi_{1,\text{EUR/JPY}}^{I} \geq 1 + \zeta_{\text{MM, EUR/JPY}}$, Market Maker \#1 adjusts its dealing price $z_{1,\text{EUR/JPY}}$ to the mid price $m_{\text{EUR/JPY}}$ (i.e., the mid point between the best quotes in (a)). This action neutralizes the existing triangular arbitrage opportunity as the new best ask quote $a_{\text{EUR/JPY}}$ (red dashed line) matches or exceeds the implied best bid quote $b_{\text{USD/JPY}}\times b_{\text{EUR/USD}}$ (green solid line).}
    \label{pvpdefense}
\end{figure}
\subsection{An additional price-entangling mechanism}

The law of one price states that in frictionless markets the prices of two assets with the same cash flows must be identical\cite{sercu2009international}. The law of one price is maintained by two distinct mechanisms, triangular arbitrage and 
and \emph{shopping around}, which promptly correct temporary gaps between the prices of two identical assets \cite{sercu2009international}. The former has been extensively described in Section \ref{sec:triangularbitrage} and it is the only way to enforce the law of one price in the standard version of the Arbitrager Model, see Section \ref{sec:Model}. The latter mechanism relates to the fact that rational traders, having detected two assets with identical cash flows but different prices, always buy the one with lower price and sell the one with higher price. This alters the demand and supply in the markets in which these assets are exchanged, thus closing the gap between their prices \cite{sercu2009international}.
Reproducing the shopping around mechanism in the Arbitrager Model requires market makers to operate in multiple LOBs. To avoid a complete overhaul of the fundamentals of the Arbitrager Model, we opt instead for a simpler stylized mechanism which retains the basic feature that distinguishes shopping around from triangular arbitrage, that is, the absence of a \emph{round trip} (e.g. JPY $\rightarrow$ EUR $\rightarrow$ USD $\rightarrow$ JPY) \cite{sercu2009international}. We assume that market makers operating in the EUR/JPY market peg their bid and ask quotes to the implied best bid and ask prices with constant probability $\gamma$, thus rejecting the dealing price update imposed by Eq.~\eqref{midpriceeq2}. For instance, the quotes of the $i$-th market maker that decides to peg its prices to the implied best quotes at time $t$ are
\begin{align}
    b_{i,\text{EUR/JPY}}(t) = b_{\text{EUR/USD}}(t) \times b_{\text{USD/JPY}}(t),\\
    a_{i,\text{EUR/JPY}}(t) = a_{\text{EUR/USD}}(t) \times a_{\text{USD/JPY}}(t).
\end{align}
This introduces an additional, simplistic mechanism through which the price of EUR/JPY is pushed towards its implied FX cross rate EUR/USD$\times$USD/JPY. 
\subsection{Cross-correlation functions and discussion}
\begin{figure}[H]
  \begin{center}
    \includegraphics[width=12.5cm]{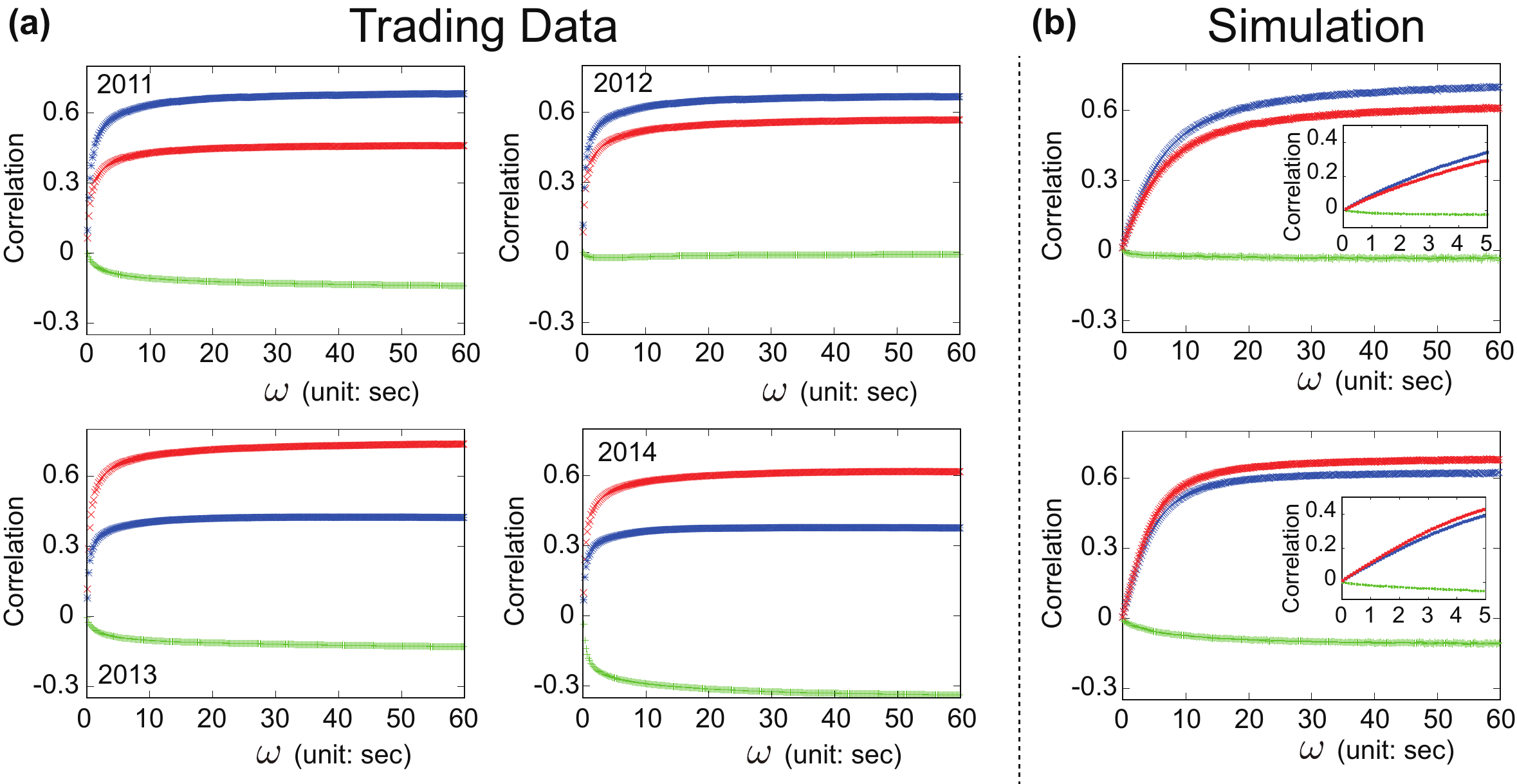}
    \caption{\textbf{Trading data vs. model based cross-correlation functions.} Cross-correlation function $\rho_{i,j}(\omega)$ for $\Delta$USD/JPY vs. $\Delta$EUR/USD (green), $\Delta$EUR/USD vs. $\Delta$EUR/JPY (blue) and $\Delta$USD/JPY vs. $\Delta$EUR/JPY (red) as a function of the time-scale $\omega$ of the underlying time series. (a) Real market data (EBS) across four distinct years (2011-2014). (b) Extended Arbitrager Model simulations. The number of participating market makers $(N_{\text{EUR/USD}},N_{\text{USD/JPY}},N_{\text{EUR/JPY}})$ are $(30,33,20)$ in the first experiment, see (b) top panel, and $(30,27,20)$ in the second experiment, see (b) bottom panel. Details on the other settings of the simulations are provided in Fig.~\ref{fig:TwoCorr}. The risk profile of the arbitrager is $\lambda_{\text{A}} = 0.01$ while the risk profiles of market makers are  $\lambda_{\text{MM, USD/JPY}} = \lambda_{\text{MM, EUR/USD}} = \lambda_{\text{MM, EUR/JPY}} = 0.001$. The pegging probability in the EUR/JPY market is $\gamma = 0.01$. The insets in (b) provide an enlarged visualization of the cross-correlations functions $\rho_{i,j}(\omega)$ on very short time-scales (i.e. $\omega < 5$ sec). Details on the initialization of the model and the conversion between simulation time (i.e., time steps) and real time (i.e., sec) are provided in Section \ref{appendix:calibration}.} 
    \label{fig:TwoCorr2}
  \end{center}
\end{figure}

Fig.~\ref{fig:TwoCorr2} reveals how the inclusion of additional features of real FX markets improves the replication of the characteristic shape of $\rho_{i,j}(\omega)$. In particular, both the data-based and model-based cross-correlation functions $\rho_{i,j}(\omega)$ approach zero on extremely short time-scales (i.e., $\omega \rightarrow 0$ sec), see insets of Fig.~\ref{fig:TwoCorr2}(b). Furthermore, the model-based $\rho_{i,j}(\omega)$ flattens on much shorter time-scales when compared to the standard Arbitrager Model, see Fig.~\ref{fig:TwoCorr}(b). This rapid stabilization is indeed observed in cross-correlation functions derived from real trading data, see Fig.~\ref{fig:TwoCorr2}(a). 
These results suggest that the discrepancies between model-based and data-based cross-correlation functions stem from the extreme simplicity of the Arbitrager Model which neglects several features and practices of real FX markets. Nonetheless, the standard Arbitrager Model succeeds in providing a comprehensive and intriguing explanation on how the dynamics of different FX rates are entangled at a microscopic level. This result is remarkable, considering the limited number of input parameters and straightforward settings that characterize our model. \\
The effort of extending the Arbitrager Model leaves us with few important insights. First, the inclusion of reacting market makers corrects the behavior of $\rho_{i,j}(\omega)$ when $\omega \rightarrow 0$ sec, that is, the model-based cross-correlation function collapses to zero as in real trading data. The key difference between arbitragers and market makers reactions to triangular arbitrage opportunities is that the former prompt simultaneous transactions, causing mid price changes in each market, while the latter cause a sequence of asynchronous mid price changes and eventually transactions. The characteristic shape of $\rho_{i,j}(\omega)$ presented in Fig.~\ref{fig:TwoCorr2}(b) is based on a set of risk profile parameters that gives market makers a predominant role at the expense of the arbitrager. This means that a large fraction of triangular arbitrage opportunities are neutralized by market makers before the arbitrager can place predatory market orders. This result cannot inform us on the fraction of opportunities that are destroyed by market makers or exploited by arbitragers in real FX markets. However, it suggests that the entanglement of the dynamics of FX rates starts at different times in each market, depending on the current state of the LOB.\\
The second insight emerging from Fig.~\ref{fig:TwoCorr2} is that the interdependencies among currencies stem from the interplay of several agents' behaviors. While the interactions between triangular arbitrage and trend-following strategies retain a primary, necessary role in the entanglement of FX rates dynamics, the introduction of a second, complementary mechanism (i.e., \emph{shopping around}) to close the gap between real and implied prices allows the model based $\rho_{i,j}(\omega)$ to stabilize on shorter time-scales $\omega$, obtaining a characteristic shape that is strongly compatible with the same function derived from real trading data. This suggests that in real FX markets additional strategies are likely to interact with triangular arbitrage and trend-following behaviors to shape the features of cross-currency correlations.
\section{The relationship between price trends and market states}
\begin{figure}[H]
\centering
\includegraphics[width=8cm]{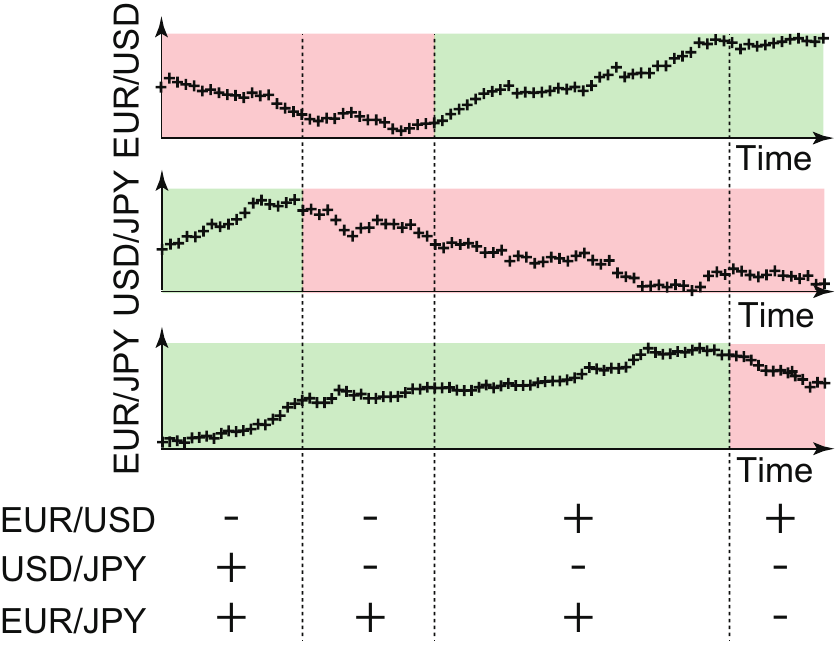}
\caption{\textbf{Price trend signs and market states.} Simulated price patterns of EUR/USD (top), USD/JPY (mid) and EUR/JPY (bottom). Periods of negative (positive) price trends are denoted by a red (green) background. Vertical dashed lines mark a change in the ecology configuration $q(t)$. Price trends $\phi_{n,\ell}$ are calculated over the most recent $n = 15$ changes in the transaction price and with scaling constant $\xi = 5$. The table below the panels combines the market states to show how the ecology configuration $q(t)$ evolves in time.}
\end{figure}
\section{Statistical properties of ecology configurations}\label{appendix:confstats}
\begin{figure}[H]
  \begin{center}
    \includegraphics[width=13cm]{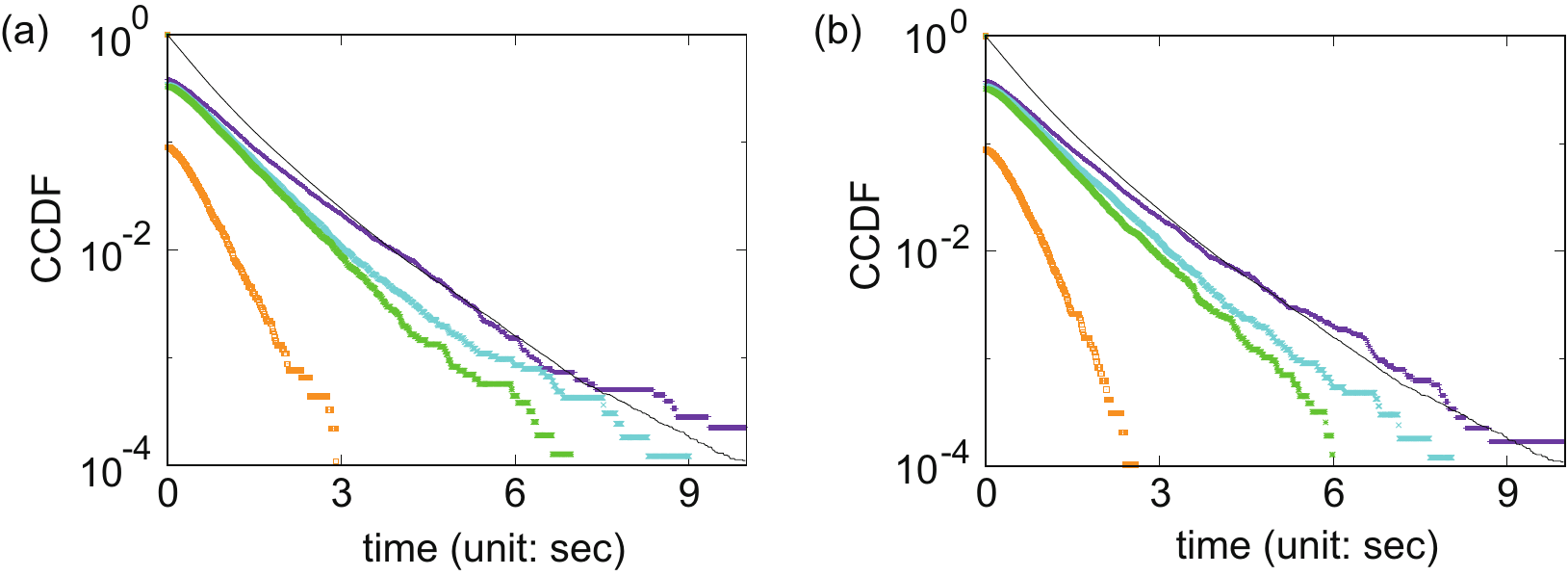}
    \caption{\textbf{Complementary cumulative distribution function (CCDF) of the time between the emergence of the first triangular arbitrage opportunity and the transition to another configuration.} The CCDFs are presented in two separate panels and each color represents a given configuration: (a) $\{+,+,+\}$ (violet), $\{-,+,+\}$ (cyan), $\{+,-,+\}$ (green) and $\{-,-,+\}$ (orange). (b) $\{-,-,-\}$ (violet), $\{+,-,-\}$ (cyan), $\{-,+,-\}$ (green) and $\{+,+,-\}$ (orange). The black lines mark the CCDF of the interval (in sec) between a random point in time and the transition to another configuration. The y-axis is visualized in the logarithmic scale. Configurations exhibit different tails of the distribution, suggesting that the probability of observing large waiting times between the emergence of the first triangular arbitrage opportunity and the transition to another configuration depends on the current combinations of market states.} 
    \label{fig:postarbitragecdf}
  \end{center}
\end{figure}
\begin{figure}[H]
  \begin{center}
    \includegraphics[width=13cm]{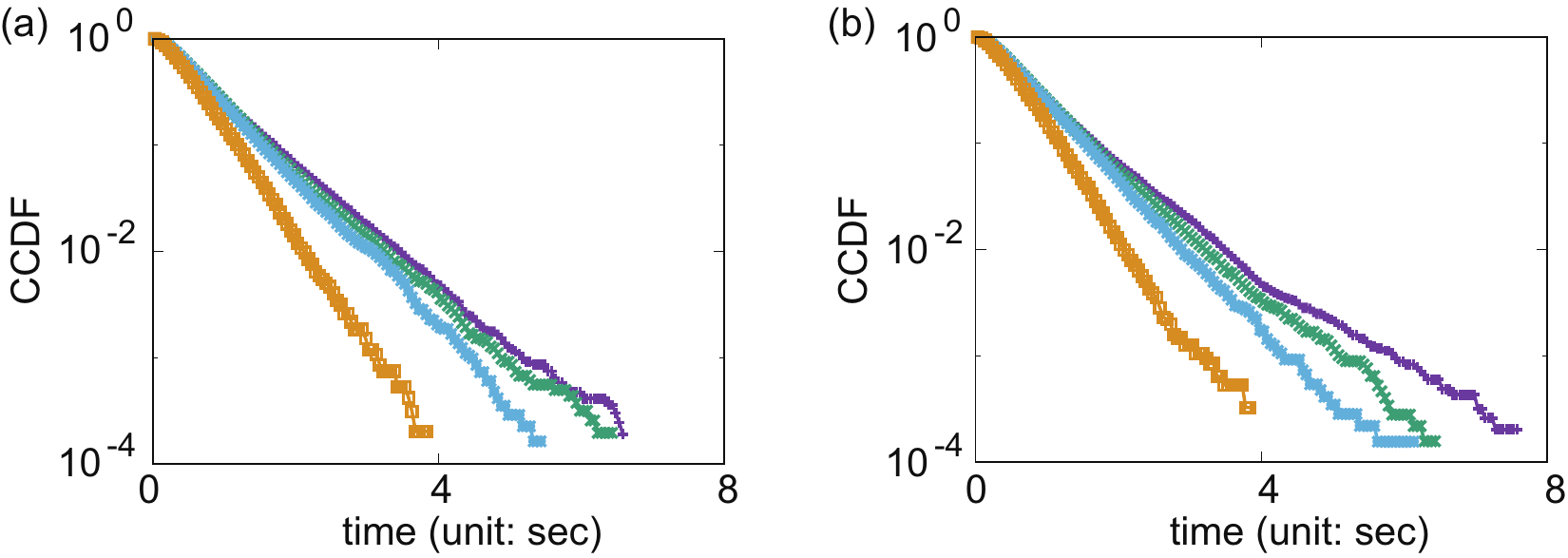}
    \caption{\textbf{Complementary cumulative distribution function (CCDF) of the time required for the first triangular arbitrage opportunity to emerge.}
    The CCDFs are presented in two separate panels and each color represents a given configuration: (a) $\{+,+,+\}$ (violet), $\{+,+,+\}$ (cyan), $\{+,-,+\}$ (green) and $\{-,-,+\}$ (orange). (b) $\{-,-,-\}$ (violet), $\{+,-,-\}$ (cyan), $\{-,+,-\}$ (green) and $\{+,+,-\}$ (orange). The y-axis is visualized in the logarithmic scale. Configurations exhibit different tails of the distribution, suggesting that the probability of observing large waiting times between the inception of the configuration and the emergence of the first triangular arbitrage opportunity depends on the current combinations of market states.} 
    \label{fig:prearbitrageccdf}
  \end{center}
\end{figure}
\begin{table}[H]
 \centering
 \caption{\textbf{Transition rates between two configurations}}
\resizebox{\textwidth}{!}{%
\begin{tabular}{|c|c|c|c|c|c|c|c|c|}
\hline
\textbf{Configuration} & \textbf{\{+, +, +\}} & \textbf{\{-, +, +\}} & \textbf{\{+, -, +\}} & \textbf{\{-, -, +\}} & \textbf{\{+, +, -\}} & \textbf{\{-, +, -\}} & \textbf{\{+, -, -\}} & \textbf{\{-, -, -\}} \\ \hline
\textbf{\{+, +, +\}} & \cellcolor[HTML]{C0C0C0} & \cellcolor[HTML]{C0C0C0}{0.358} & \cellcolor[HTML]{C0C0C0}{0.336} & 0.010 & 0.227 & 0.028 & 0.026 & 0.014 \\ \hline
\textbf{\{-, +, +\}} & \cellcolor[HTML]{C0C0C0}{0.375} & \cellcolor[HTML]{C0C0C0} & \cellcolor[HTML]{C0C0C0}0.027 & {\color[HTML]{000000} 0.218} & 0.011 & 0.329 & 0.014 & 0.026 \\ \hline
\textbf{\{+, -, +\}} & \cellcolor[HTML]{C0C0C0}{0.364} & \cellcolor[HTML]{C0C0C0}0.026 & \cellcolor[HTML]{C0C0C0} & {\color[HTML]{000000} 0.217} & 0.012 & 0.014 & 0.341 & 0.026 \\ \hline
\textbf{\{-, -, +\}} & 0.035 & {\color[HTML]{000000} 0.302} & {\color[HTML]{000000} 0.295} &  & 0.006 & 0.028 & 0.030 & {\color[HTML]{000000} 0.305} \\ \hline
\textbf{\{+, +, -\}} & {\color[HTML]{000000} 0.304} & 0.030 & 0.028 & 0.007 &  & {\color[HTML]{000000} 0.294} & {\color[HTML]{000000} 0.302} & 0.035 \\ \hline
\textbf{\{-, +, -\}} & 0.026 & 0.340 & 0.014 & 0.012 & {\color[HTML]{000000} 0.220} & \cellcolor[HTML]{C0C0C0} & \cellcolor[HTML]{C0C0C0}0.026 & \cellcolor[HTML]{C0C0C0}{0.363} \\ \hline
\textbf{\{+, -, -\}} & 0.028 & 0.013 & 0.330 & 0.011 & {\color[HTML]{000000} 0.217} & \cellcolor[HTML]{C0C0C0}0.027 & \cellcolor[HTML]{C0C0C0} & \cellcolor[HTML]{C0C0C0}{0.374} \\ \hline
\textbf{\{-, -, -\}} & 0.015 & 0.027 & 0.027 & 0.226 & 0.010 & \cellcolor[HTML]{C0C0C0}{0.335} & \cellcolor[HTML]{C0C0C0}{0.359} & \cellcolor[HTML]{C0C0C0} \\ \hline
\end{tabular}%
}
\caption*{Rows (Columns) indicate the departed (reached) configuration. We count how many times the system transitioned between two specific configurations and normalize this number by the total number of transitions from the departed configuration. The two grey portions of the matrix mark the first (upper-left) and second (lower-right) clusters discussed in Section \ref{sec:results2}.}
\label{tab:tranmatrix}
\end{table}
\begin{figure}[H]
\centering
\includegraphics[width=8cm]{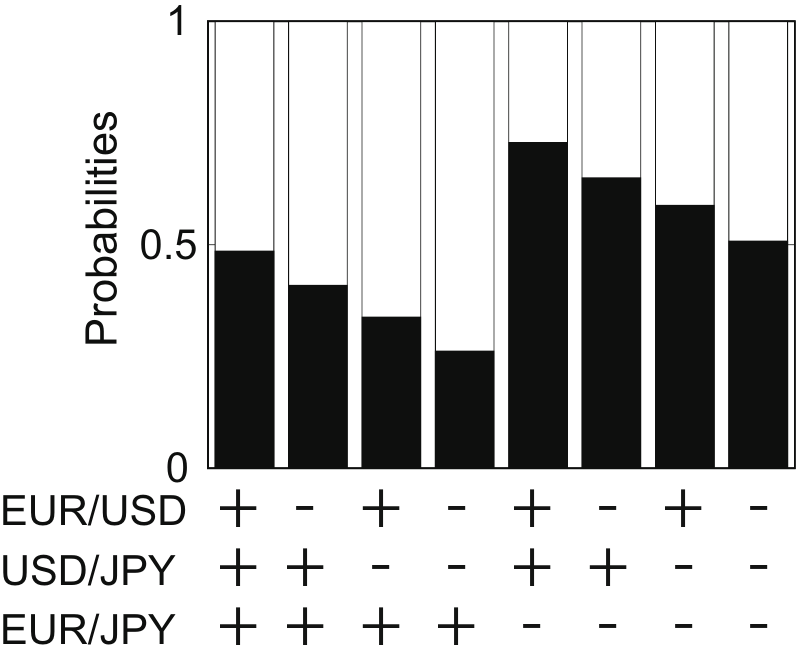}
    \caption{\textbf{Fraction of triangular arbitrage opportunities of the first and second type in each ecology configuration.} Black bars denote the incidence of type 1 opportunities, see Eq.~\eqref{mu1}, while white bars represent the incidence of type 2 opportunities, see Eq.~\eqref{mu2}. We notice that one type appears more frequently than the other, depending on the considered configuration.}
    \label{fig:mubarplot}
\end{figure}
\begin{figure}[H]
\centering
\includegraphics[width=8cm]{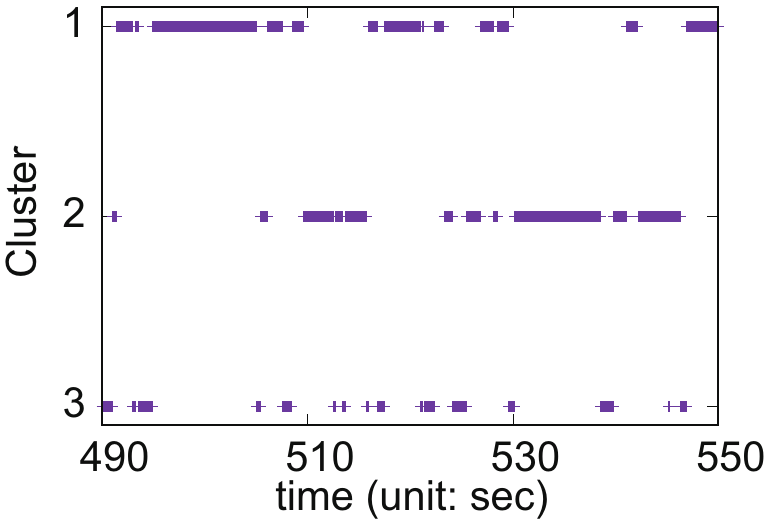}
    \caption{\textbf{The sequence of transitions between configurations exhibits a clustered behavior.} The x-axis represents an arbitrary time window of the experiment. The y-axis splits the eight ecology configurations in three groups - from top to bottom: (1) cluster 1, (2) cluster 2 and (3) $\{-,-,+\}$ and $\{+,+,-\}$, which are the two configurations that do not belong to any cluster. Details on the concept of configuration clustering are provided in Section \ref{sec:results2}.
    At each point in time we identify the current ecology configuration and add a marker to the group it belongs to. A visual inspection of the figure reveals the presence of time windows in which the system moves between configurations belonging to the same cluster, corresponding to the long, uninterrupted lines observed in groups 1 and 2, but not in group 3. These peculiar dynamics favor the appearance of configurations belonging to groups 1 and 2 at the expenses of those belonging to group 3, see Fig.~\ref{fig:statusDuration}(b).  Details on the conversion between simulation time (i.e., time steps) and real time (i.e., sec) are provided in Section \ref{appendix:calibration}.}
    \label{fig:clusteranalysis}
\end{figure}

\begin{figure}[H]
\centering
\includegraphics[width=8cm]{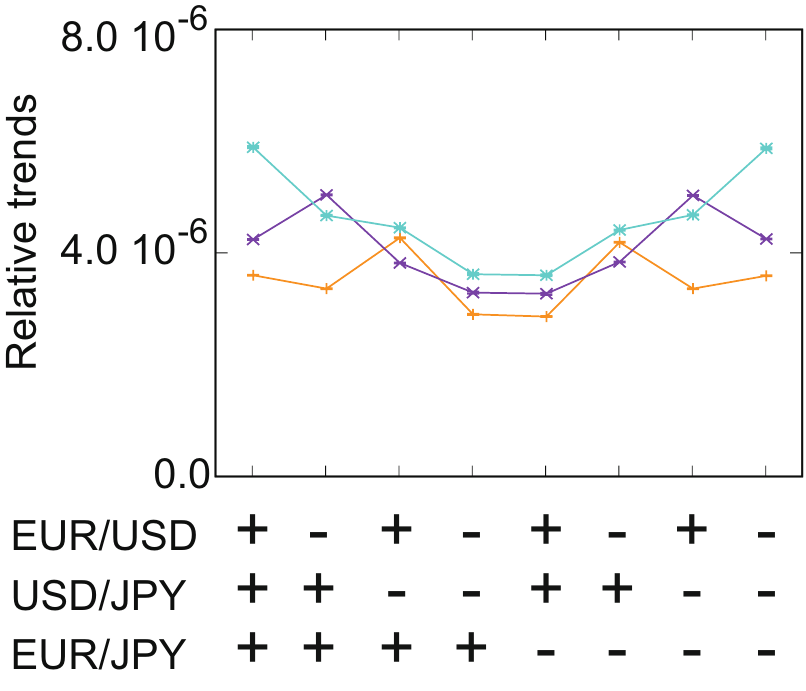}
    \caption{\textbf{Price trends and changes in market states.} Sample averages of the normalized absolute value of the price trend $\langle|\phi_{n,\ell}(t)|/p_{\ell}(t_0)\rangle$ for EUR/USD (orange), USD/JPY (violet) and EUR/JPY (cyan). Normalizing by the initial center of mass $p_{\ell}(t_0)$ allows us to compare the price trends across markets with different price magnitudes. We exclusively sample the value $|\phi_{n,\ell}(t)|/p_{\ell}(t_0)$ at the emergence of each triangular arbitrage opportunity and consider each configuration independently. As arbitrager's market orders alter price trends, the value of $|\phi_{n,\ell}(t)|/p_{\ell}(t_0)$, where $t$ is the time step when $\mu^{I}$ or $\mu^{II}$ exceeds the unit, informs us on how currently hard is to flip the state of the $\ell$-th market. For instance, we consider $\{+,+,+\}$ and observe that $\langle|\phi_{n,\ell}(t)|/p_{\ell}(t_0)\rangle$ is much higher in EUR/JPY than in EUR/USD and USD/JPY. This is reflected in the probabilities of transitioning from $\{+,+,+\}$ to other configurations. Flipping EUR/JPY before the other two markets, causing a transition to $\{+,+,-\}$, occurs in 22.7\% of the cases. However, flipping EUR/USD or USD/JPY first, causing a transition to $\{-,+,+\}$ or $\{+,-,+\}$, occur in 35.8\% and 33.6\% of the cases respectively, see Table~\ref{tab:tranmatrix}.
    }
    \label{fig:trendstrengthhist}
\end{figure}

\end{document}